\documentclass[11pt,a4paper]{article}
\usepackage{jheppub}

\usepackage{epsfig,multicol,bbm}
\usepackage{axodraw4j}
\usepackage{pstricks}
\usepackage{color}

\newcommand{\gsim}{ \mathop{}_{\textstyle \sim}^{\textstyle >} }
\newcommand{\lsim}{ \mathop{}_{\textstyle \sim}^{\textstyle <}}

\newcommand\fverb{\setbox\fverbbox=\hbox\bgroup\verb}
\newcommand\fverbdo{\egroup\medskip\noindent%
			\fbox{\unhbox\fverbbox}\ }
\newcommand\fverbit{\egroup\item[\fbox{\unhbox\fverbbox}]}
\newbox\fverbbox

\newcommand{\beq}{\begin{eqnarray}}
\newcommand{\eeq}{\end{eqnarray}}

\newcommand{\be}{\begin{eqnarray}}
\newcommand{\ee}{\end{eqnarray}}

\newcommand{\mev}{{\rm MeV}}

\newcommand{\tev}{{\rm TeV}}
\newcommand{\gev}{{\rm GeV}}
\newcommand{\kev}{{\rm keV}}
\newcommand{\Mev}{{\rm MeV}}
\newcommand{\kms}{{\rm km/s}}

\title{Low Energy INTEGRAL Positrons from eXciting Dark Matter}
\author[a]{Rob Morris}
\author[a,b]{Neal Weiner}
\affiliation[a]{Center for Cosmology and Particle Physics\\ Department of Physics, New York University\\ New York, NY 10003, USA}
\affiliation[b]{School of Natural Sciences\\ Institute for Advanced Study\\Princeton, NJ 08540, USA}
\emailAdd{Rob.Morris@physics.nyu.edu}\emailAdd{neal.weiner@nyu.edu}

\abstract{The origin of the $e^+e^-$ 511 keV line observed by INTEGRAL remains unclear. The rate and morphology of the signal have prompted questions as to whether dark matter could play a role. We explore the case of dark matter upscattering in the framework of eXciting Dark Matter (XDM), where WIMPs $\chi$, interacting through a new dark force, scatter into excited states $\chi^*$, which subsequently emit $e^+e^-$ pairs when they de-excite. We numerically compute the cross sections for two Yukawa-coupled DM particles upscattering into excited states, specifically considering variations motivated by recent N-body simulations with additional baryonic physics. We find that that $l>0$ components of the partial-wave decomposition are often significant contributions to the total cross section and that for reasonable ranges of parameters dark matter can produce the $\sim 10^{43} \:e^+/{\rm s}$ observed by INTEGRAL.}


\begin{document} 
\maketitle
\section{Introduction}
Over the past several years there has been increasing evidence for a variety of astrophysical anomalies. These anomalies have generally taken the form of the presence of a new signal of radiation or cosmic rays, beyond what was conventionally expected. They come in the form of high energy $e^+e^-$ sources, as seen by PAMELA \cite{Boezio:2008mp} and Fermi \cite{Abdo:2009zk}, microwave emission from the galactic center \cite{Finkbeiner:2003im,Finkbeiner:2004us}, and diffuse gamma rays from a broad ($20-40^\circ$) range around the galactic center \cite{Finkbeiner:2003im}. All of these anomalies can be related to the presence of a new, primary source of high energy ($\sim 100\:\gev$) $e^+e^-$, which may be attributable to a weak-scale dark matter origin, either through annihilation or decay. 

An outlier in the list of astrophysical anomalies is the INTEGRAL 511 $\kev$ signal \cite{Attie:2003, Knodlseder:2003sv} (see \cite{Prantzos:2010wi} for a recent discussion). Both bulge and disk-correlated sources are observed. The morphology of the bulge component is best modeled by a combination of gaussians with widths $3^\circ$ and $11^\circ$.  The bulge and disks fluxes are comparable at roughly $10^{-3}\: {\rm ph}\: {\rm cm}^{-2}\: {\rm s}^{-1}$. This corresponds to a positron production rate of $10^{43}\: e^+/{\rm s}$. While this signal has persisted over decades, originally seen in the early 1970's \cite{Johnson:1972, Johnson:1973, Haymes:1975}, the origin of the enormous source of positrons needed to explain it remains elusive. In particular, the $10^{43} \:e^+/{\rm s}$ concentrated in the galactic center region, as well as the highly spherical morphology are a challenge to achieve from most galactic sources, which tend to trace the disk. Alternative explanations, such as low mass X-ray binaries (LMXBs) \cite{Prantzos:2004}  could possibly provide a candidate \cite{Weidenspointner:2006nu}, although no point source 511 $\kev$ emission has yet been observed \cite{DeCesare:2010dc}. Likewise, it has been suggested that the transport of the positrons produced in the galactic disk into the galactic center could provide the rate \cite{Lingenfelter:2009kx}, although the precise dynamics that achieves this and yields such a spherical morphology is unclear.

At the same time, the possible connection of this signal to dark matter is even less obvious, principally because of two facts: first, that the shape of the 511 $\kev$ line is sufficiently narrow as to constrain the injection energy to be below $\sim 10\:\Mev$ \cite{Beacom:2005qv}. Second, for a weak scale dark matter candidate, the rate is orders of magnitude above what is expected from a thermal WIMP annihilation signal. Such ideas have induced people to focus on $\Mev$ scale dark matter particles \cite{Boehm:2003ha, Boehm:2003hm, Boehm:2003bt} as an alternative, but a connection to more massive theories of dark matter remains appealing.

A candidate explanation of this was the ``eXciting Dark Datter'' (XDM) proposal \cite{Finkbeiner:2007kk}. In this proposal, an excited state $\chi^*$ of the dark matter $\chi$ is postulated with a splitting $\delta \sim 2 m_e$ \footnote{For related work, see \cite{Pospelov:2007xh}.}. Dark matter - dark matter collisions mediated by a new dark force produce an excitation $\chi \rightarrow \chi^*$, followed by the decay $\chi^*\rightarrow \chi e^+e^-$. The appealing aspect of this idea is that one converts the {\em kinetic} energy of a WIMP into positrons. Since this is a scattering, rather than annihilation process, the cross section can be much larger, giving the possibility of yielding the enormous $O(10^{43} \:e^+/\mathrm{s})$ observed in the galactic center.

Such a proposal is not without problems however. In order to produce the large rates, a large cross section $\sigma \sim 1/q^2$ was needed, forcing the inclusion of a new GeV-scale mediator, $\phi$, and even then remains challenging \cite{Pospelov:2007xh, Chen:2009dm, Chen:2009av}. Intriguingly, because XDM freezes out by annihilating into $\phi$, the annihilation $\chi \chi \rightarrow \phi \phi$ will naturally produce a hard positron signal \cite{Finkbeiner:2007kk, Cholis:2008vb}. This simple idea led to the proposal of a ``unified'' model \cite{TODM}, which simultaneously addresses PAMELA/Fermi, WMAP, INTEGRAL and DAMA (through the inelastic dark matter scenario \cite{Smith:2001hy}). It was argued that in such models $\phi$ can mediate a Sommerfeld enhancement of the annihilation\cite{TODM,Pospelov:2008jd} \footnote{The Sommerfeld enhancement was first discussed in the context of dark matter by \cite{Hisano:2003ec, Hisano:2004ds}.} to give the rates observed at PAMELA.

Our focus here is to reconsider more carefully the low-energy positron signal. Because the signal involves a non-perturbative process, mediated by multiple $\phi$ exchanges, calculating the expected rates can have important subtleties. Moreover, the possible rates depend sensitively on both astrophysical and particle physics parameters. By approaching this in detail, we hope to understand what the natural expectation for a rate in the galactic center would be and what ranges of astrophysical and model parameters can explain the observed INTEGRAL signal.

\subsection{Models of XDM and Signals at INTEGRAL}
Models of XDM are simple to construct \cite{Finkbeiner:2007kk}. The first proposed model involved a dark matter as a complex scalar $\chi$ coupled to a new gauge boson $\phi_\mu$ of a dark gauge group $U(1)_d$, with a small $\lsim \rm GeV$ mass. We assume that after $U(1)_d$ breaking a small splitting $\delta$ arises between the real scalar states. 
\be
{\cal L} = ( D_\mu \chi_i )^* D_\mu \chi_i + \frac{1}{4}F^d_{\mu\nu} F^{d\mu\nu}+\epsilon F_{\mu\nu} F^{d\mu\nu}+m^2 \phi_{ \mu} \phi^{\mu}+M^2_i  \chi_i^* \chi_i+M_i \delta_i \chi_i \chi_i + {\rm h.c.}
\label{eq:lagrangianscalar}
\ee
One can also replace the scalar easily with a pseudo-Dirac fermion,
\be
{\cal L} = i \bar \chi_i \not \! \! D \chi_i + \frac{1}{4}F^d_{\mu\nu} F^{d\mu\nu}+\epsilon F_{\mu\nu} F^{d\mu\nu}+m^2 \phi_{ \mu} \phi^{\mu}+M_i \bar \chi_i \chi_i+\delta_i  \chi_i \chi_i +{\rm h.c.}
\label{eq:lagrangian}
\ee
The relic abundance is established by freezing out through annihilations into $\phi$, which yields the usual ``WIMP miracle'' result that for $\left<\sigma v\right> \approx 3 \times 10^{-26} {\rm cm}^2$, $\Omega h^2 \approx 0.1$. 

\begin{figure}[t]
a)\scalebox{0.7}{\fcolorbox{white}{white}{
  \begin{picture}(304,240) (115,-91)
    \SetWidth{1.0}
    \SetColor{Black}
    \Line[arrow,arrowpos=0.5,arrowlength=5,arrowwidth=2,arrowinset=0.2](112,22)(256,22)
    \Line[arrow,arrowpos=0.5,arrowlength=5,arrowwidth=2,arrowinset=0.2](256,22)(368,54)
    \Photon(256,22)(256,-58){7.5}{4}
    \Line[arrow,arrowpos=0.5,arrowlength=5,arrowwidth=2,arrowinset=0.2](112,-58)(256,-58)
    \Line[arrow,arrowpos=0.5,arrowlength=5,arrowwidth=2,arrowinset=0.2](256,-58)(368,-90)
    \Text(96,28)[lb]{\Large{\Black{$\chi$}}}
    \Text(96,-68)[lb]{\Large{\Black{$\chi$}}}
    \Text(230,-20)[lb]{\Large{\Black{$\phi$}}}
    \Text(352,28)[lb]{\Large{\Black{$\chi^*$}}}
    \Text(352,-68)[lb]{\Large{\Black{$\chi^*$}}}
  \end{picture}
}}
b)\scalebox{0.7}{\fcolorbox{white}{white}{
  \begin{picture}(288,240) (115,-91)
    \SetWidth{1.0}
    \SetColor{Black}
    \Line[arrow,arrowpos=0.5,arrowlength=5,arrowwidth=2,arrowinset=0.2](128,48)(256,48)
    \Line[arrow,arrowpos=0.5,arrowlength=5,arrowwidth=2,arrowinset=0.2](256,48)(368,128)
    \Photon(256,48)(288,-16){7.5}{4}
    \Line[arrow,arrowpos=0.5,arrowlength=5,arrowwidth=2,arrowinset=0.2](288,-80)(288,-16)
    \Line[arrow,arrowpos=0.5,arrowlength=5,arrowwidth=2,arrowinset=0.2](288,-16)(352,-16)
    \Text(112,64)[lb]{\Large{\Black{$\chi^*$}}}
    \Text(320,128)[lb]{\Large{\Black{$\chi$}}}
    \Text(250,0)[lb]{\Large{\Black{$\phi^*$}}}
    \Text(272,-96)[lb]{\Large{\Black{$e^+$}}}
    \Text(368,-16)[lb]{\Large{\Black{$e^-$}}}
  \end{picture}
}}
\caption{Process responsible for the INTEGRAL signal in the XDM model with a vector mediator. {\it (left)} Scattering excitation process for XDM is enhanced by multiple $\phi$ exchange (not shown). {\it (right)} The excited states decays back to the ground state through an offshell $\phi$, producing $e^+e^-$ pairs.}
\end{figure}
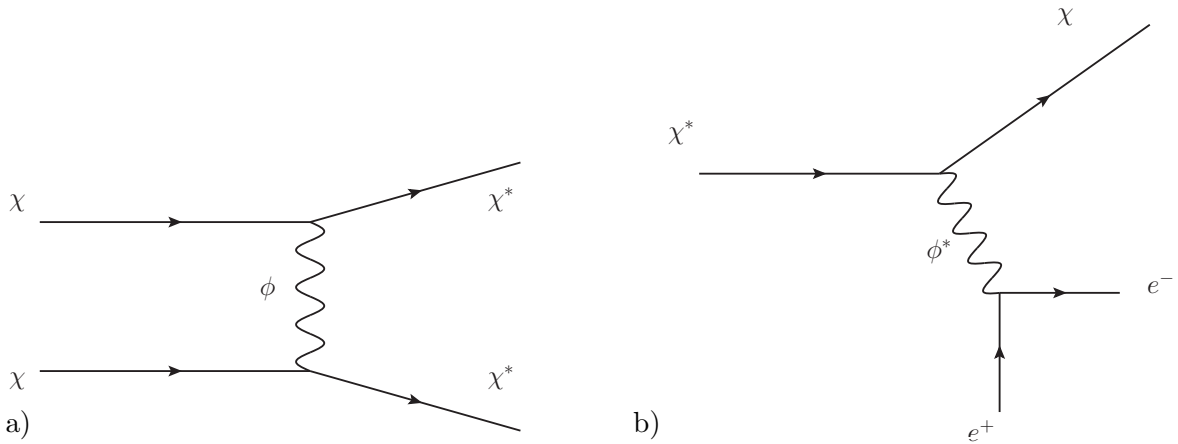

The light force carrier couples to SM states arises through kinetic mixing, as first described by \cite{Holdom:1985ag}. Although this mixing can be quite small when only addressing the INTEGRAL signal, for sizeable $\epsilon$, terrestrial experiments (such as fixed target, beam dump and searches at low-energy accelerators) can provide limits on these light bosons \cite{Reece:2009un,Bjorken:2009mm,Batell:2009di}. Indeed, there are already important new limits \cite{Andreas:2010tp,Archilli:2011nh,Merkel:2011ze,  Abrahamyan:2011gv}  and additional proposals for new searches \cite{Proposal,Essig:2010xa,HPS,Wojtsekhowski:2009vz,Freytsis:2009bh}.

Even with this setup and a light mediator, it is not obvious that such a model can actually achieve such a large rate. We can estimate this by using the Einasto profile \cite{Merritt:2005xc} with parameters set by the A-1 run of the Aquarius simulation \cite{Springel:2008cc}. In the presence of a light mediator, a natural scale for the scattering rate is set by the geometric cross section $\sigma \sim \pi/q^2$. At threshold, $q^2 =  m_\chi \delta$, which we take as a lower bound on the natural scale of the scattering. Assuming a relative velocity of $2 \times 10^{-3}c$, $m_\chi \approx 1\:\tev$ and $\delta \approx 1\:\Mev$, one estimates a rate in the inner 2 kpc of $7\times10^{42} \:e^+/\mathrm{s}$.

This can be increased, for instance for particles well above threshold, where the momentum transfer is low. Our order of magnitude calculation, however, assumes that {\em all} particles are kinematically capable of scattering, which is an overestimate. For the U(1) vector interaction, both WIMPs are excited, requiring $2\delta$ of available energy. For this to occur, the characteristic velocity of a $1\:\tev$ WIMP must be $\gsim 425\:\kms$. If the velocity dispersion is low $\sim \sqrt{3/2}\times 220\:\kms$ -- comparable to most estimates of the local value -- only a small fraction of particles will be kinematically capable of scattering. Indeed, with a low and constant velocity dispersion, it is essentially impossible to achieve these high rates \cite{Pospelov:2007xh}. Such observations prompted the development of scalar mediated models \cite{Finkbeiner:2007kk} and non-Abelian models \cite{TODM,ArkaniHamed:2008qp,Baumgart:2009tn,Chen:2009av}, as well as models with metastable states \cite{Chen:2009av,Finkbeiner:2009mi,Cirelli:2010nh,Cline:2010kv}, where the threshold is lower.

However, it would be surprising if the velocity dispersion would stay constant, and a number of recent simulations with baryons \cite{RomanoDiaz:2008wz,Governato:2007,2010MNRAS.407..435A,Pedrosa:2009rw, Tissera:2009cm} see an increase of the dispersion roughly as a power law of the radius as one moves towards the galactic center. If this is true, then in the inner 1 kpc the majority of particles would be capable of scattering and the high velocities can allow for larger cross sections and scattering rates. Even so, it is not clear that such high rates can be achieved, with \cite{Chen:2009av} finding no points in parameter space that can achieve these high rates.

In this paper, we will re-examine this question. In section \ref{sec:calapproach} we will discuss our approach to calculating the scattering process, which is consistent with previous approaches using a partial wave analysis. In section \ref{sec:rates} we convert these partial wave amplitudes into the expected rates for INTEGRAL and explore the contributions from each partial wave mode. Using Einasto parameters from the Aquarius A-1 DM-only simulation \cite{Navarro:2008kc} we find rates of $10^{41}$--$10^{42} \:e^+/\mathrm{s}$. In this section we also explore variations of individual profile parameters and find that they can change the rates by a factor of 5--10. In section \ref{sec:TWvariations} we consider more recent simulations including baryons \cite{Pedrosa:2009rw, Tissera:2009cm} and find rates of $10^{42}$--$10^{43} \:e^+/\mathrm{s}$, enough to explain the excess.  Here we also compare our work with previous work on this matter. Finally, in section \ref{sec:conclusions}, we discuss connections to other signals and conclude.

\section{Calculational Approach}
\label{sec:calapproach}
Let us consider a two-state system where the states are separated by a mass splitting $\delta$. We are interested in $2 \to 2$ scattering where two particles enter in the ground state and both are upscattered into the excited state. In this scenario, the total splitting between the incoming two-particle wavefunction and the outgoing two-particle wavefunction is defined to be $\Delta$. To avoid confusion we will only refer to this total splitting, $\Delta$. Note that this total splitting between the two-particle wavefunctions due to a double excitation, where $\Delta=2\delta$, is equivalent to a single excitation with $\Delta=\delta$.

Because the particles are moving non-relativistically, the system is simply governed by the Schr\"odinger equation, which we will solve in the basis of partial waves. We assume the particles are attracted by a Yukawa-type force mediated by a particle with mass $m_\phi$. We will consider XDM-type scenarios where the coupling is off-diagonal. Note that we use $\alpha_d$ for the fine structure constant so as not to confuse it with the Einasto profile parameter $\alpha$. For each partial wave mode $l$, the reduced Schr\"odinger equation has the form

\begin{equation}
\frac{1}{m_\chi}
\left(
\begin{array}{c}
\chi_1''(x) \\ 
\chi_2''(x)
\end{array}
\right )
=
V
\cdot
\left(
\begin{array}{c}
\chi_1(x) \\ 
\chi_2(x)
\end{array}
\right )
+
\left(\frac{l(l+1)}{m_\chi r^2}-E\right)
\left(
\begin{array}{c}
\chi_1(x) \\ 
\chi_2(x)
\end{array}
\right )
\end{equation}
where $E$ is the energy of the two-state system and the potential $V$ is given by
\begin{equation}
V = \left(
\begin{array}{cc}
0 & -\alpha_d \frac{e^{-m_\phi r}}{r}\\ 
-\alpha_d \frac{e^{-m_\phi r}}{r} & \Delta 
\end{array}
\right)
\end{equation}

Following \cite{TODM} we restate the Schr\"odinger equation into the dimensionless parameters $\epsilon_v \equiv v/\alpha_d$, $\epsilon_\delta \equiv \sqrt{\Delta / m_\chi}/\alpha_d$ and $\epsilon_\phi \equiv m_\phi / (\alpha_d m_\chi)$. Using this reparameterization (and rescaling $r$ by $r \rightarrow \alpha_d m_\chi r$) we are left with
\begin{eqnarray}
\left(
\begin{array}{c}
\chi_1''(x) \\ 
\chi_2''(x)
\end{array}
\right )
=
\left(
\begin{array}{cc}
\frac{l(l+1)}{r^2}-\epsilon_{v}^{2} & -{\rm e}^{-\epsilon_{\phi}r} \\ 
-{\rm e}^{-\epsilon_{\phi}r} & \frac{l(l+1)}{r^2}+\epsilon_{\delta}^{2}-\epsilon_{v}^{2}
\end{array}
\right)
\cdot
\left(
\begin{array}{c}
\chi_1(x) \\ \chi_2(x)
\end{array}
\right )
\end{eqnarray}

For boundary conditions, we chose the wave functions to be regular at the origin. Remember that for the reduced Schr\"odinger equation: $\chi(r) = r R(r)$, where $R(r)$ is the radial part of the solution to the full (spherically symmetric) Schr\"odinger equation. This gives the two conditions $\chi_1(0)=0$ and $\chi_2(0)=0$. We then impose that $\chi_2$ is composed of purely outgoing spherical waves at infinity. This leaves us with one condition left and to set it we simply normalize $\chi_1 = 1$ at infinity. We are free to do this because we will always be concerned with ratios of the wavefunctions. 

We would like to note two things about these expressions.  The first is that we can see this parameterization is equivalent to that of \cite{Chen:2009dm} by noting that $\Gamma = 1/\epsilon_\delta^2$, $\Upsilon = (\epsilon_v/\epsilon_\delta^2)^2$ and $\eta = (\epsilon_\phi/\epsilon_\delta^2)$. The second is that the equations depend only on $v$, $\alpha_d$ and the ratios $\Delta/m_\chi$ and $m_\phi/m_\chi$. In this work we will only consider $\alpha_d = 1/100$ and total splittings of $1\:\Mev$ and $2\:\Mev$. Also, since we are concerned only with thermalized cross sections, the velocity will always be integrated over. This leaves us with only two physical parameters to scan over: the WIMP mass and the force carrier mass.

We would also like to reiterate the statement of \cite{Slatyer:2009vg} that much of the parameter space is numerically unstable and thus we found it difficult to accurately compute partial wave modes higher than $l=7$. We found the most stable computational method to be a variant of the shooting method called the chasing method \cite{Berezin:1965,Na:1979} with roughly 50 digits of precision during the internal computation. We imposed strict error tests on the auxiliary system for the unknown boundary value and rejected any data points that failed those tests. In section \ref{sec:partialwaves} we will present the results of these computations as plots of the partial wave modes. In section \ref{sec:rateconvergence} we discuss the convergence of our sums over partial waves, but we note that if anything this technique underestimates the rates by truncating the sum at $l=7$.

\subsection{Partial Waves}
\label{sec:partialwaves}

We are ultimately interested in the rate of $e^+ e^-$ pair production, which depends on the thermalized scattering cross section.  But to compute the cross sections for upscattering, we must first compute the partial wave amplitudes, $f_l$. For a given $\alpha_d$, $\Delta/m_\chi$ and $m_\phi/m_\chi$, each $f_l$ is a function of $\epsilon_v$. We find that values of $\epsilon_v$ greater than about $0.35$ are highly suppressed, independent of the WIMP characteristics. This constraint comes from assuming a maximum escape velocity of $1000\:\kms$ in the center of the galaxy. As discussed in section \ref{sec:crosssections}, the low end of the velocity integral is set by the threshold velocity $v_{th} = 2 \sqrt{\Delta/m_\chi}$. Thus we need only compute the $f_l$ functions in this window.

We define the partial wave amplitudes as
\begin{equation}
f_l = \frac{k'|\chi_{2\mbox{\textsc{\tiny\em out}}}|^{2}}{k|\chi_{1\mbox{\textsc{\tiny\em in}}}|^{2}}
\end{equation}
where $k'$ is the momentum of the excited state and $k$ is the momentum of the ground state. This is the same definition used in \cite{Chen:2009dm}. It can be obtained by using the conservation of probability flux to define the scattering amplitude. We numerically separate the wave functions into incoming and outgoing components by taking Fourier transforms.

To get a sense of these partial wave amplitudes we present a selection of them as functions of $\epsilon_v$ for various masses and splittings. They have similar shapes to the partial wave amplitude functions found by \cite{Chen:2009av}. Any gaps along the functions where the plot marker is missing correspond to a point that yielded some sort of numerical error.  
\begin{figure}[h]
\centering
\includegraphics[width=\textwidth]{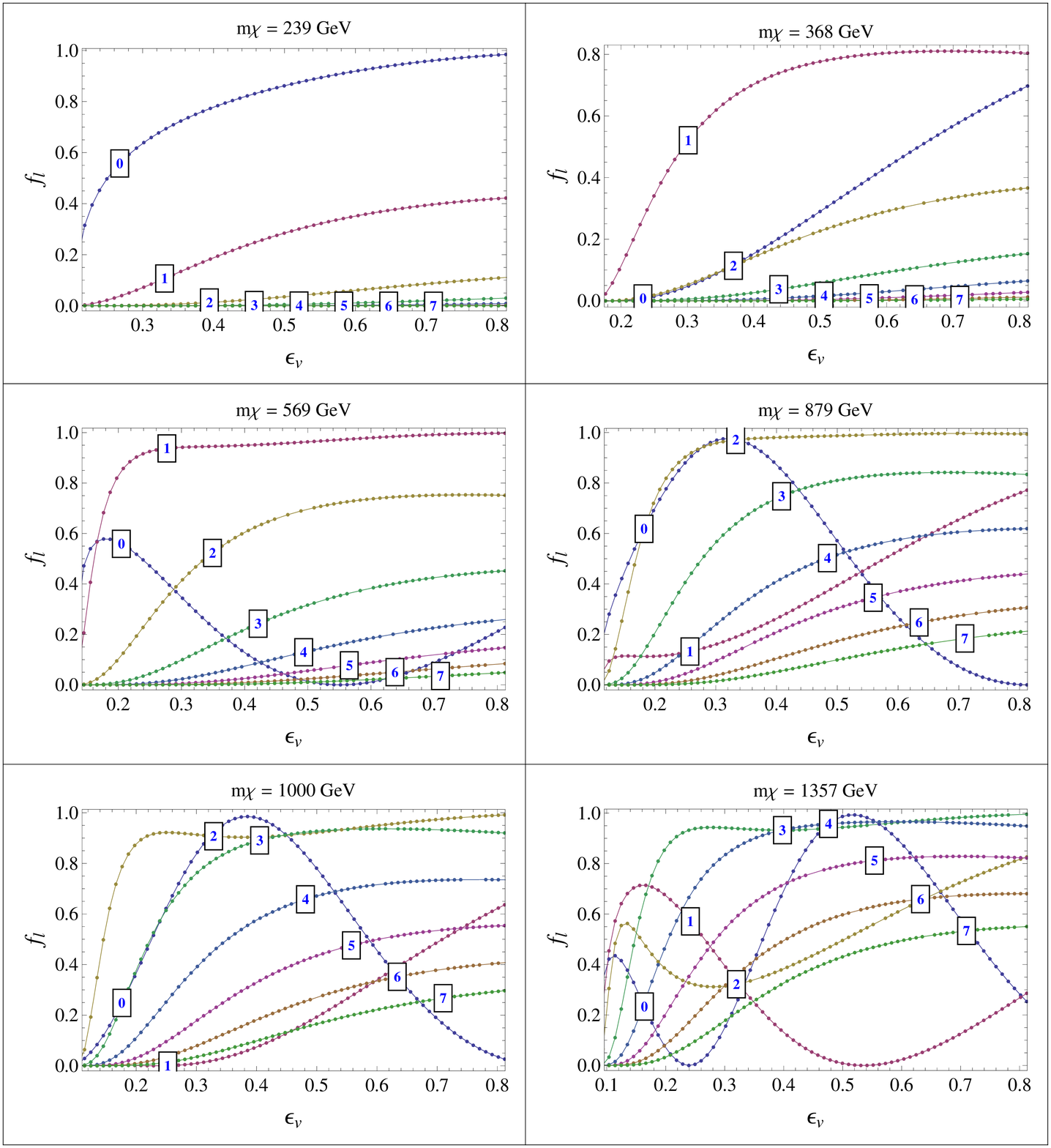}
\caption{Partial wave amplitudes as functions of $\epsilon_v$ for various $m_\chi$. $m_\phi=1\:\gev$ and $\Delta=1\:\Mev$}
\end{figure}
\begin{figure}[h]
\centering
\includegraphics[width=\textwidth]{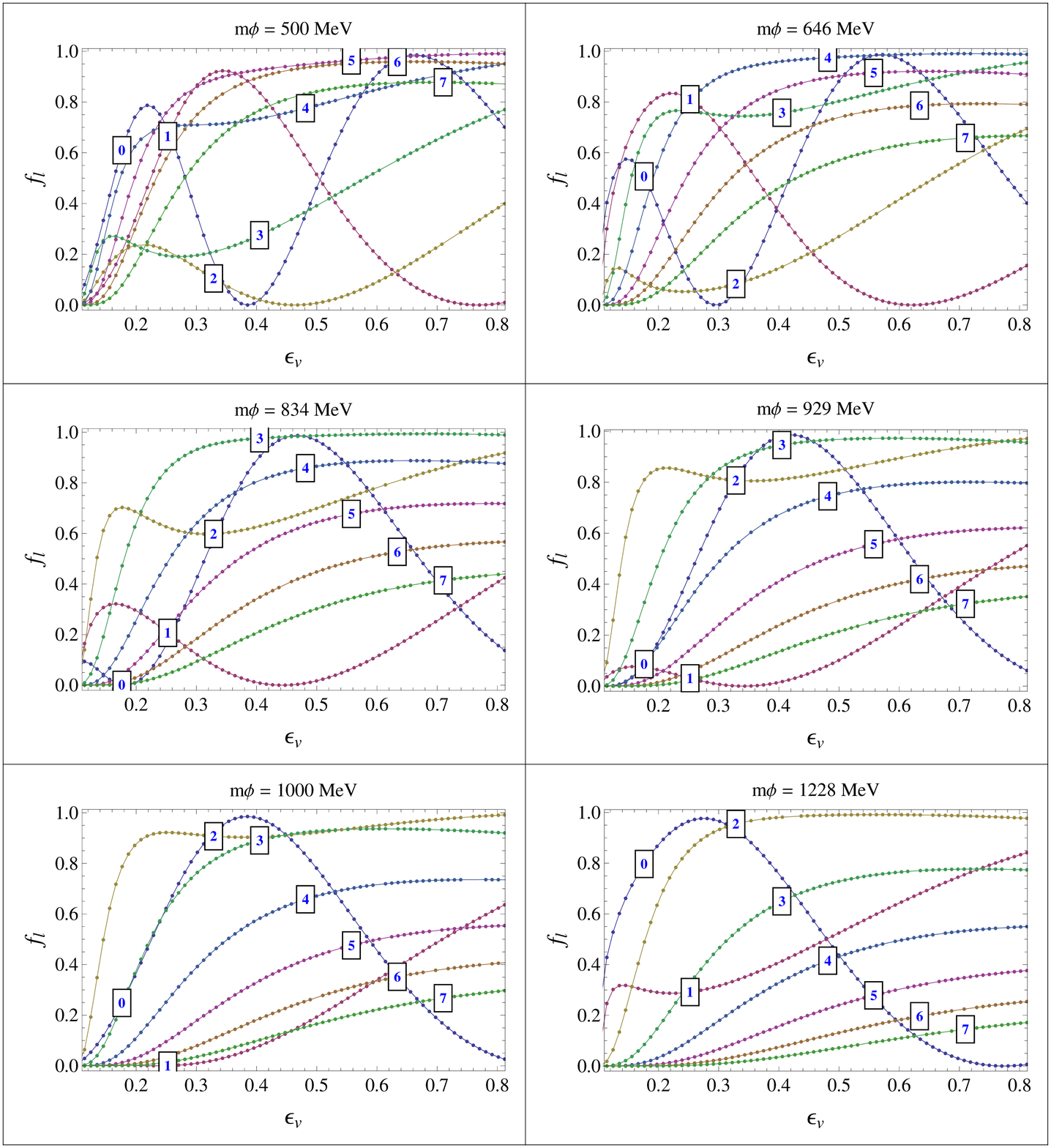}
\caption{Partial wave amplitudes as functions of $\epsilon_v$ for various $m_\phi$. $m_\chi=1\:\tev$ and $\Delta=1\:\Mev$}
\end{figure}
While the partial wave amplitudes we show here have similar shapes to those of \cite{Chen:2009av}, the parameter space considered here does not completely overlap with the parameter ranges considered there. We will return to this point and its implications in section \ref{sec:clinecompare}.

\subsection{Velocity Profile}
\label{sec:velocityprofile}
Now with the partial wave amplitudes in hand, we are nearly ready to construct the cross sections. But because the partial wave amplitudes are velocity dependent, our cross sections will also depend on velocity. We then must thermally average our cross sections over the WIMP velocity distribution in the galaxy, in order to yield accurate results. We will first then develop the appropriate velocity distribution before we can calculate the thermally-averaged cross sections.

In the rest frame of the galaxy, we assume the WIMPs to have at every point a Maxwell-Boltzmann speed distribution peaked around the RMS speed $v_0$. In principle $v_0$ could be a function of galactic radius. As stated above, the constant $v_0$ case doesn't yield the desired rates, so we will consider the case where $v_0(r) = (220\:\kms)(\frac{r}{8\:\mathrm{kpc}})^{-1/4}$. In section \ref{sec:TWvariations} we will consider velocity dispersions specific to a set of DM simulations including baryons as well.  The WIMP speed distribution is truncated by the escape velocity, $v_{esc}$. The escape velocity is assumed to be a function of galactic radius. 

To estimate the escape velocity profile, we assume the rotational velocity profile is fairly flat as a function of radius \cite{Fich:1989bs}. The condition for uniform circular motion then gives us
\begin{equation}
\label{eq:circmotion}
\frac{v_c^2}{r}=\frac{G M(r)}{r^2}
\end{equation}
which we can then plug into the escape velocity condition
\begin{equation}
\frac{1}{2}v_{esc}^2=\int_r^\infty \frac{G M(\tilde{r})}{\tilde{r}^2}\,d\tilde{r}.
\end{equation}
We can break the integral up into an integral from $r$ to $R_\odot$ and an integral from $R_\odot$ to $\infty$. The second integral is just our local escape velocity. Plugging in for $M(r)$ from equation \ref{eq:circmotion} and assuming a local escape velocity of $600\:\kms$ \cite{Smith:2006ym} we are left with the escape velocity distribution
\begin{equation}
v_{esc}^2=2 v_c^2 \ln \left(\frac{R_\odot}{r} \right) +(600\:\kms)^2.
\end{equation}
This allows us to know the escape velocity in terms of the circular velocity $v_c$ with a reasonable assumption of the total (dark+baryonic) mass distribution function for the galaxy.

We will work in the center of momentum frame and will thus need to transform the two particles' three-dimensional velocity distributions into a one-dimensional relative velocity distribution. To do this, we first change variables from the two particles' velocities ($\vec{v_1}$ and $\vec{v_2}$) to the total and relative velocities: $\vec{v_t} = \vec{v_1} + \vec{v_2}$ and $\vec{v_r} = \vec{v_1} - \vec{v_2}$. We can then integrate out the angular parts and radial part of the total velocity with the conditions that $\vec{v_1}$, $\vec{v_2} < v_{esc}$. This leaves us (after proper normalization) with a velocity distribution that is solely a function of $v_r$:
\begin{equation}
V(v_r)=
   \frac{e^{-\frac{v_r^2}{2 v_0^2}} v_r \left(\sqrt{\pi } v_r
   e^{\frac{4 v_{esc}^2+v_r^2}{4 v_0^2}}
   \mathrm{erf}(-\frac{v_r-2 v_{esc}}{2 v_0})+2
   v_0 \left(e^{\frac{v_r^2}{2 v_0^2}}-e^{\frac{v_{esc}v_r}{v_0^2}}
   \right)\right) \theta \left(2
   v_{esc}-v_r\right)}{v_0 \left(\sqrt{2 \pi } v_0
   e^{\frac{v_{esc}^2}{2 v_0^2}}
   \mathrm{erf}(\frac{v_{esc}}{\sqrt{2} v_0})-2
   v_{esc}\right)^2}
\end{equation}
which has been normalized by two factors of $N$, where 
\begin{equation}
N = \int_{0}^{v_{esc}}4\pi v^{2} e^{-v^{2}/(2 v_{0}^{2})}\,dv.
\end{equation}

\begin{figure}
\includegraphics[width=.5\textwidth]{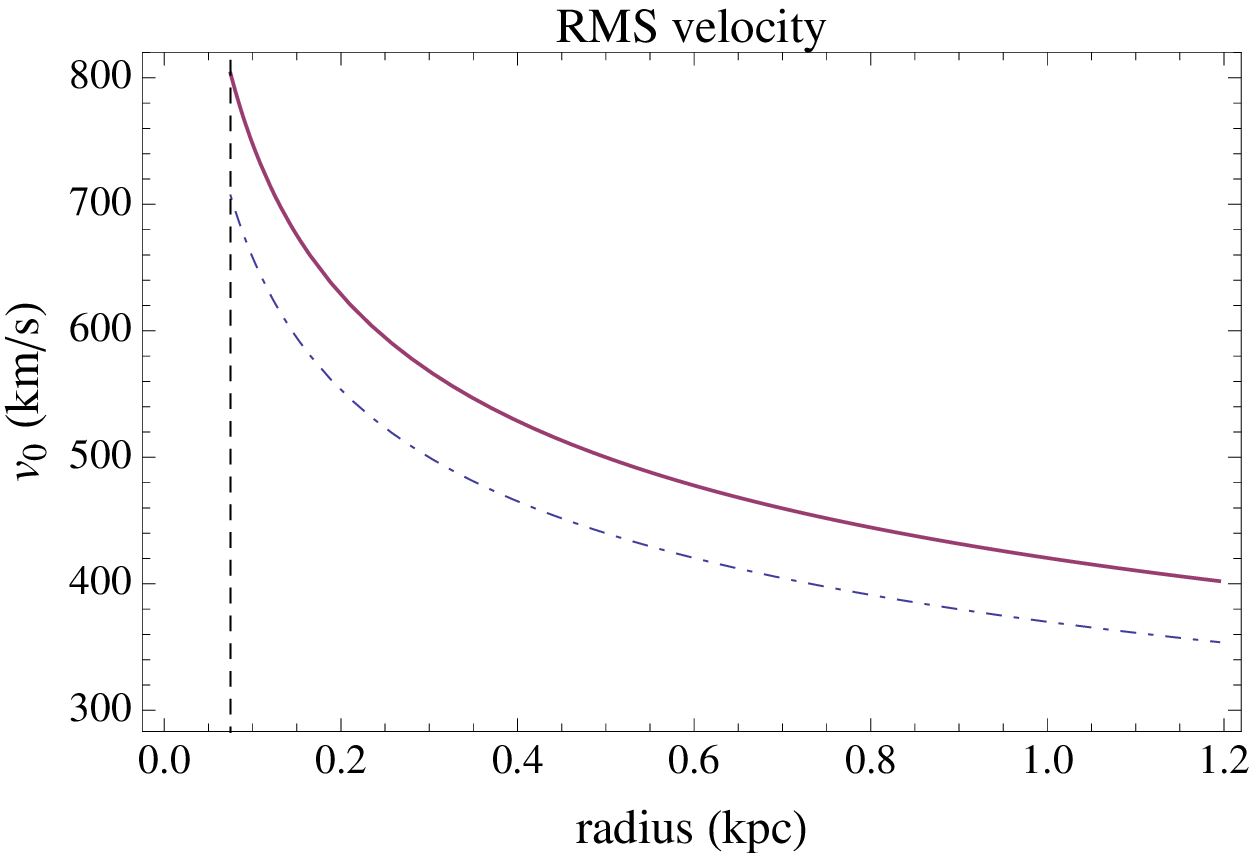}
\includegraphics[width=.5\textwidth]{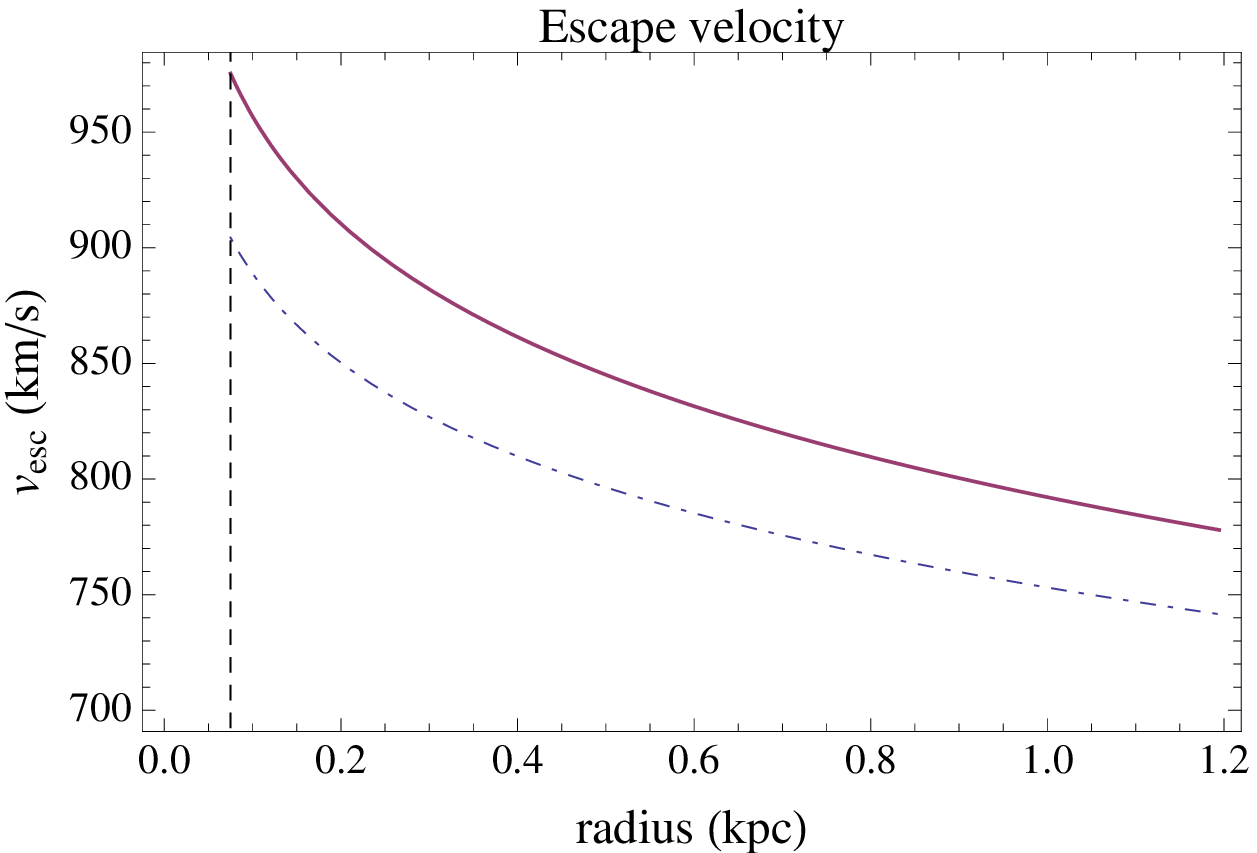}
\caption{{\it (left)} RMS velocity profile, {\it (right)} escape velocity profile. {\it (dot-dashed)} $v_c=220\:\kms$, {\it (solid)} $v_c=250\:\kms$.}
\label{fig:v0vescplots}
\end{figure}
\begin{figure}
\centering
\includegraphics[scale=.75]{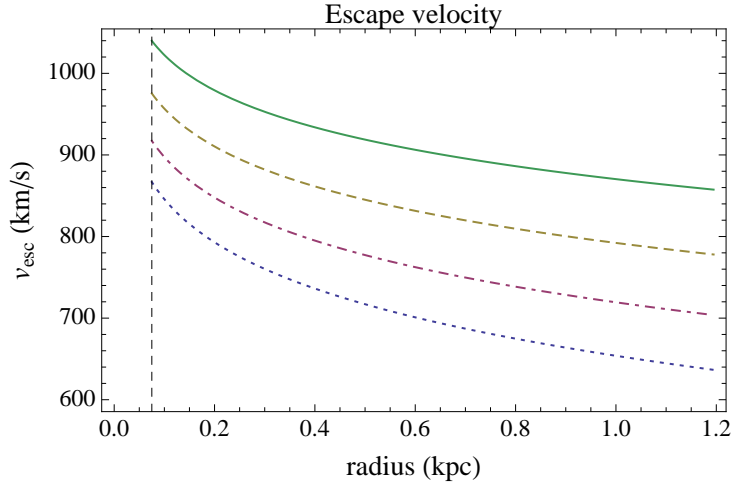}
\caption{Escape velocity profiles for different local escape velocities. {\it dotted}: $v_{loc}=400\:\kms$, {\it dot-dashed}: $v_{loc}=500\:\kms$, {\it dashed}: $v_{loc}=600\:\kms$ and {\it solid}: $v_{loc}=700\:\kms$.}
\label{fig:vescvaryinglocal}
\end{figure}

We include for reference in figure \ref{fig:v0vescplots} plots of the RMS velocity profile and the escape velocity profile for two different local circular velocities. In this work we will generally assume a local circular velocity, $v_c$, of $250\:\kms$ \cite{Fich:1989bs}. We also include in figure \ref{fig:vescvaryinglocal} the escape velocity profiles for $v_c=250\:\kms$ and various local escape velocities. We will consider the effects these escape velocity profiles have on the rates in section \ref{sec:vlocvariations}. We plot the profiles up to a radius of $r\simeq 1.2\:\rm{kpc}$, which roughly corresponds to the angular size of the INTEGRAL signal.  The dashed vertical line in the plots marks the lower limit on our integral which we have chosen to be $r=0.075\:\rm{kpc}$ (to avoid the cusp as $r\to0$).

\subsection{Cross Sections}
\label{sec:crosssections}
Armed with the partial wave amplitudes and the relative velocity distribution we are finally ready to construct the thermalized cross sections. Standard partial wave scattering theory tells us the cross section for a given $l$ mode is given by
\begin{eqnarray}
\label{eq:crosssections}
\sigma_{l}(\epsilon_v) &=& \int{d\Omega |(2l+1)\frac{S_l}{2 i k'} P_l(\cos \theta)|^2} \nonumber \\
                       &=& \frac{\pi (2l+1)}{k'^2} \frac{k'|\chi_{2\mbox{\textsc{\tiny\em out}}}|^{2}}{k|\chi_{1\mbox{\textsc{\tiny\em in}}}|^{2}}  \\
                       &=& \frac{\pi (2l+1)}{m_\chi^2 v_{r}^2}f_{l}(\epsilon_v) \nonumber
\end{eqnarray}
where $\epsilon_v$ is $v/\alpha_d = (v_r/2)/\alpha_d$ and we have used the ground state mass in the last line rather than the excited state. The thermalized cross section is then given by the sum of these $l$ partial cross sections integrated over the velocity distribution:
\begin{equation}
\label{eq:avgcrosssections}
\left< \sigma v \right>= \sum_{l=0}^\infty\int_{v_{th}}^{2 v_{esc}}\sigma_{l}(\frac{v_{r}}{2 \alpha_d}) v_r V(v_r)\,dv_{r}.
\end{equation}
Here the lower limit is given by the threshold velocity for inelastic scattering. The threshold velocity, $v_{th} = 2\sqrt{\frac{\Delta}{m_\chi}}$, is the minimum relative velocity needed to scatter when there is a mass difference and it satisfies $2\times\frac{1}{2} m_\chi (\frac{v_{th}}{2})^2 = \Delta$.

\subsection{Einasto Density Profile}
\label{sec:einasto}
The rate of scattering per volume of two identical WIMPs is given by 
\begin{equation}
\frac{d \Gamma}{d V} = \frac{1}{2}n_\chi^2\left<\sigma v\right>
\end{equation}
where $n_\chi$ is the WIMP number density and the factor of $1/2$ avoids double-counting. In general, $n_\chi$ is a function of galactic radius. Recent halo simulations including the effects of baryons \cite{Onorbe:2006qk, Gao:2007gh, Pedrosa:2009rw, Tissera:2009cm} favor Einasto density profiles so those are the ones we shall consider. 

The generic Einasto density profile is given by
\begin{equation}
\log\left(\frac{\rho(r)}{\rho_{-2}}\right) = \frac{-2}{\alpha}\left[\left(\frac{r}{r_{-2}}\right)^{\alpha}-1\right].
\end{equation}
We can eliminate $\rho_{-2}$ by assuming a local WIMP density of 0.4 GeV/cm\textsuperscript{3}. This leaves us with the density profile
\begin{equation}
\rho(r) = \frac{2}{5}\exp\left[\frac{2}{\alpha}\left(\left(\frac{R_\odot}{\,r_{-2}}\right)^\alpha-\left(\frac{r}{r_{-2}}\right)^\alpha\right)\right].
\end{equation}
\begin{figure}[h]
	\centering
		\includegraphics[scale=.75]{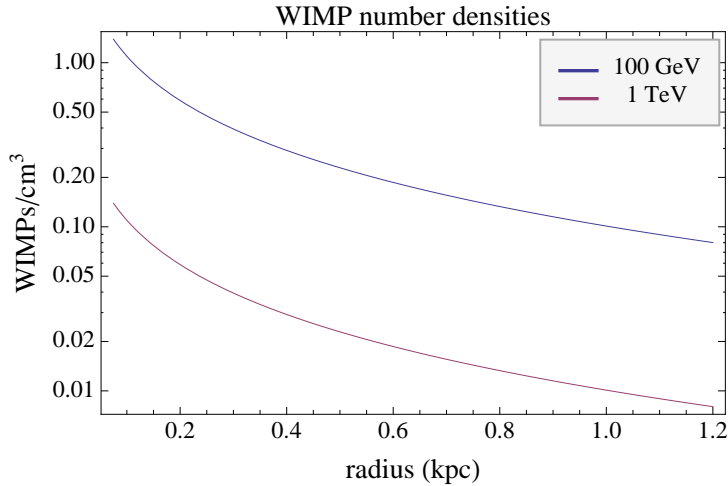}
		\caption{Number densities from the Einasto profile using $\alpha=0.17$ and $r_{-2} =15.79$.}
	\label{fig:numberdensities}
\end{figure}

Here $\alpha$ (not to be confused with the dark fine structure constant) determines the cuspiness of the profile and $r_{-2}$ is the radius at which the logarithmic slope takes the isothermal value. As a baseline, we will assume $\alpha=0.17$ and $r_{-2} = 15.79$, which correspond to the A-1 run of the Aquarius simulation \cite{Navarro:2008kc}, though we will we consider variations of these parameters in section \ref{sec:profilevariations}. As a reference, we include plots of the WIMP number density for both the $100\:\gev$ and $1\:\tev$ WIMPs for these parameters.

We can then integrate $d \Gamma/d V$ over a volume in the galaxy to get the total rate of scatterings in that volume. We chose to integrate in a small region in the center of the galaxy with radius $r_c$. Remember that both the density profile and the thermalized cross section (via the RMS and escape velocities) depend on galactic radius. The final expression for the total rate of scattering, $\Gamma$, is then
\begin{equation}
\Gamma = \frac{1}{2}\int_0^{r_c}4\pi r^2 \left(\frac{\rho(r)}{m_\chi}\right)^2 \left< \sigma v(r) \right> \, dr.
\end{equation}
Note that due to the cuspiness and uncertainty in the center of the galaxy, we do not actually integrate from 0---we start the integral at $r=0.075\:\rm{kpc}$.

\section{Rates}
\label{sec:rates}

We numerically solve the Schr\"odinger equation for many points in the $m_\chi$--$m_\phi$ parameter space to construct the partial wave amplitude functions. We first hold $m_\phi$ fixed at $1\:\gev$ and vary $m_\chi$ from $100\:\gev$ to $5\:\tev$.  Next we hold $m_\chi$ fixed at $1\:\tev$ and vary $m_\phi$ from $500\:\mev$ to $5\:\gev$.  We do this for both $\Delta=1\:\Mev$ and $\Delta=2\:\Mev$. We then numerically integrate the differential scattering rate from the center of the galaxy to 1.2\:kpc with the Einasto profile parameters $\alpha=0.17$ and $r_{-2} = 15.79$, assuming a local DM density of $0.4\:\gev/\mathrm{cm}^3$. This distance roughly corresponds to the angular width of the INTEGRAL signal. For a detailed discussion of what constraints can be inferred by requiring the dark matter to fit the angular profile of the signal, we would refer the reader to the more detailed discussions in \cite{Abidin:2010ea,Cline:2010kv}.

\begin{figure}[t!]
\includegraphics[width=.5\textwidth]{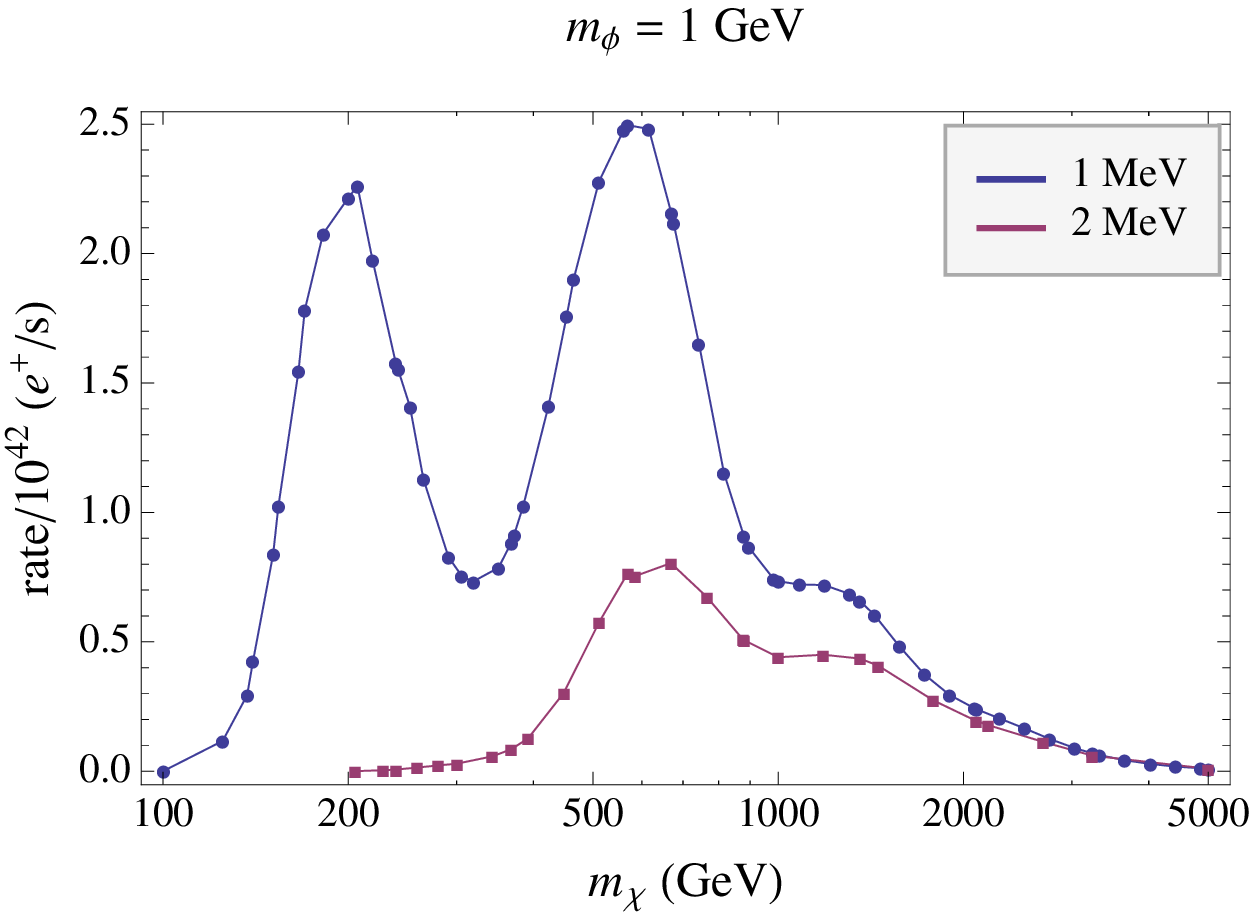}
\includegraphics[width=.5\textwidth]{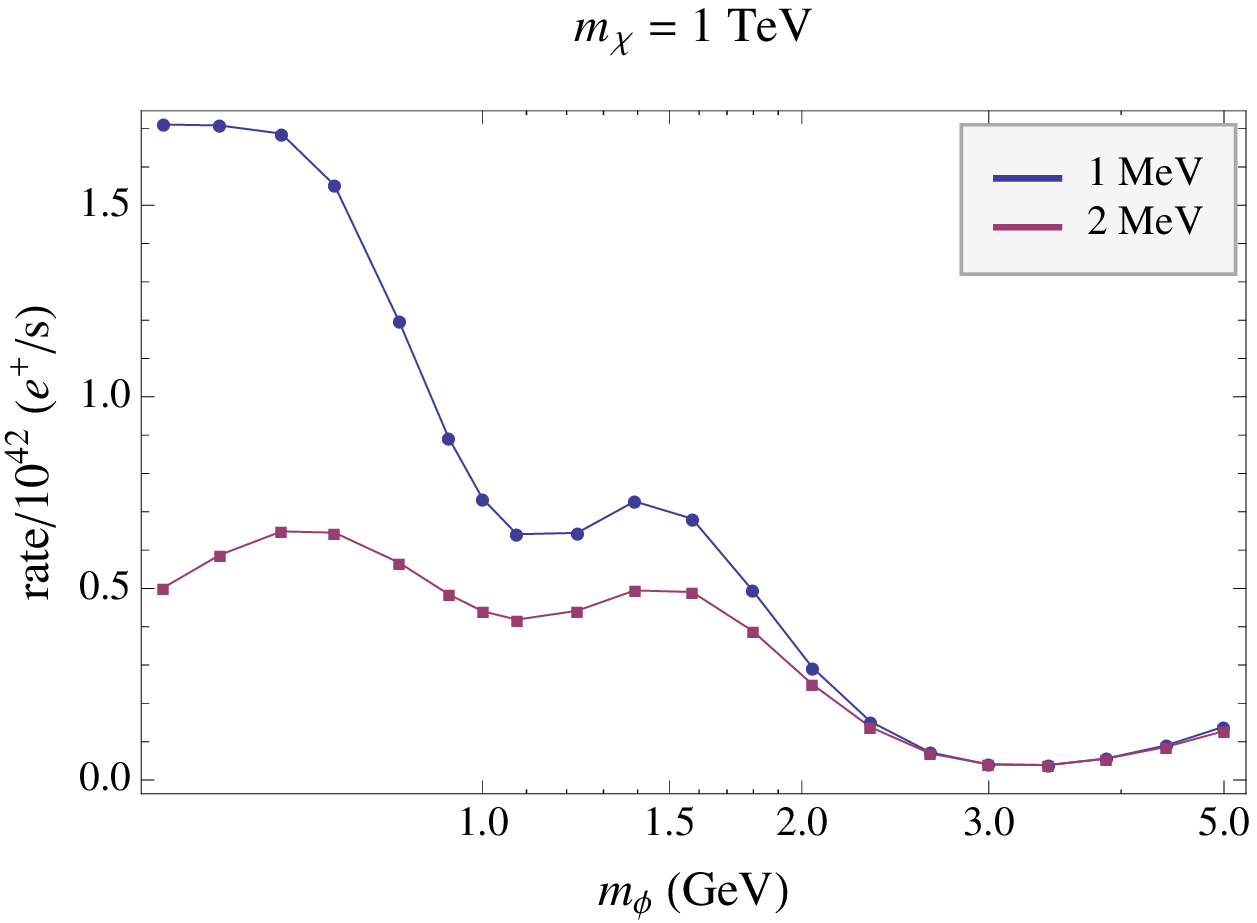}
\caption{Rates of $e^+e^-$ production using Aq-A-1 profile parameters. \it{disks}: $\Delta=1\:\Mev$, \it{squares}: $\Delta=2\:\Mev$.}
\label{fig:rates}
\end{figure}

We can see from figure \ref{fig:rates} that with these parameters, the rates are of the order $O(10^{41})-O(10^{42})$. We also see there are resonance regions along variations in $m_\chi$. There are also smaller resonances in $m_\phi$, but generally lower $m_\phi$'s give higher rates.  Of the two resonance regions along $m_\phi = 1\:\gev, \Delta = 1\:\Mev$, the one peaked around $m_\chi \approx 200\:\gev$ is probably heavily dependent on the velocity profile, as we will show in section \ref{sec:rateconvergence}. The peak around $m_\chi \approx 600\:\gev$ is more robust considering possible changes in the halo model. In section \ref{sec:profilevariations} we will show that these rates can easily go up an order of magnitude or more by varying the profile parameters.
 
\subsection{Partial Wave Contributions}
\label{sec:partialwavecontributions}

\begin{figure}[h]
\includegraphics[width=.47\textwidth]{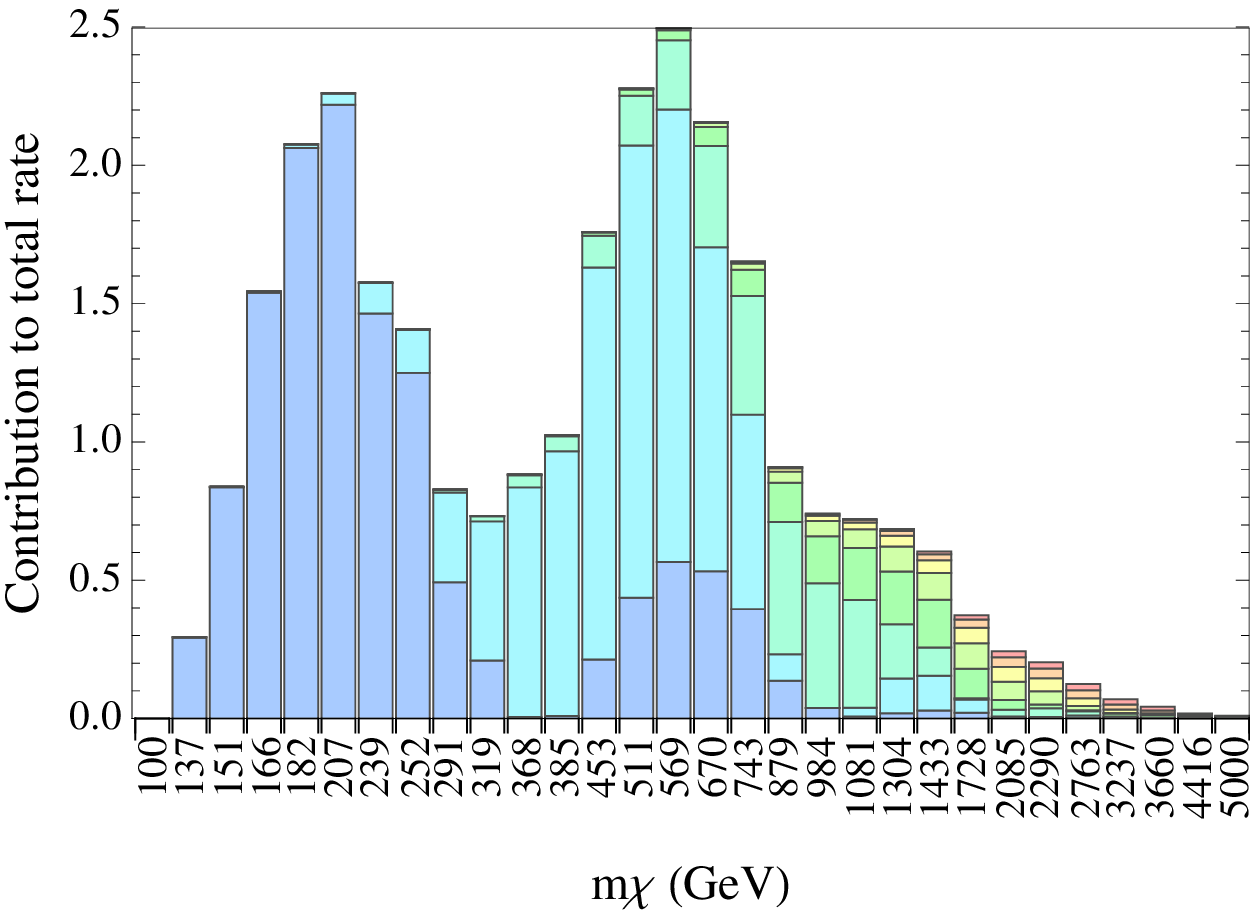}
\includegraphics[width=.53\textwidth]{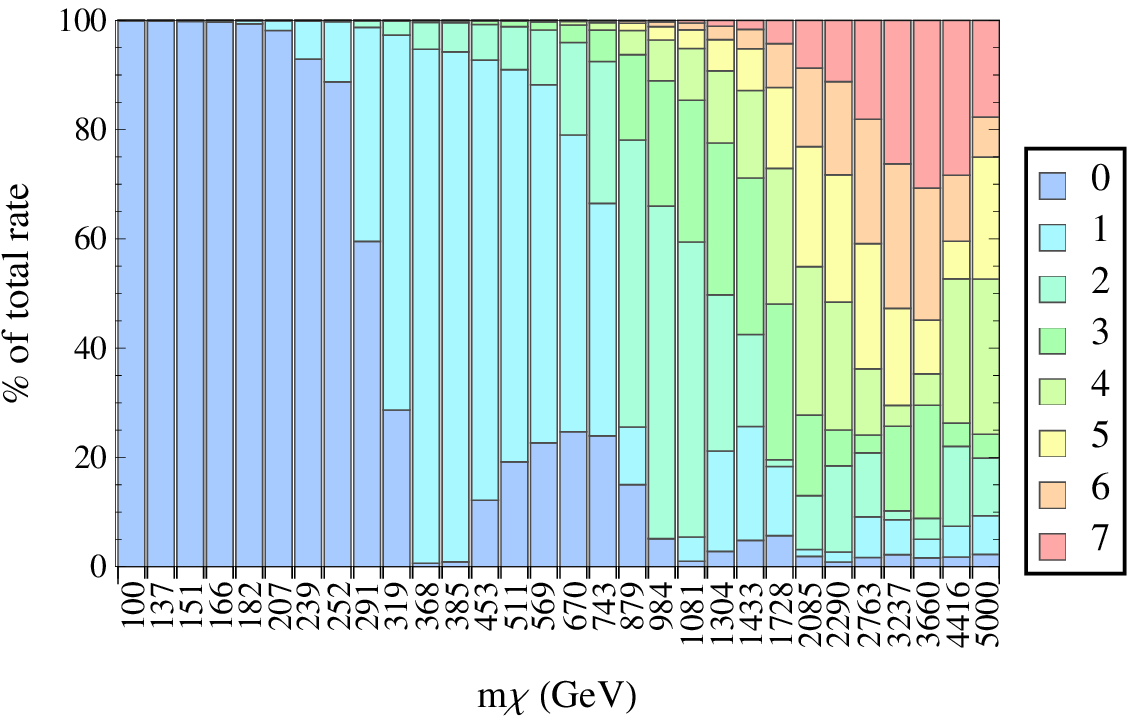}
\caption{Individual partial wave contributions for $m_\phi = 1\:\gev$ and $\Delta = 1\:\Mev$ varying $m_\chi$.}
\label{fig:pwcontributionsmchi1MeV}
\end{figure}

\begin{figure}[h]
\includegraphics[width=.47\textwidth]{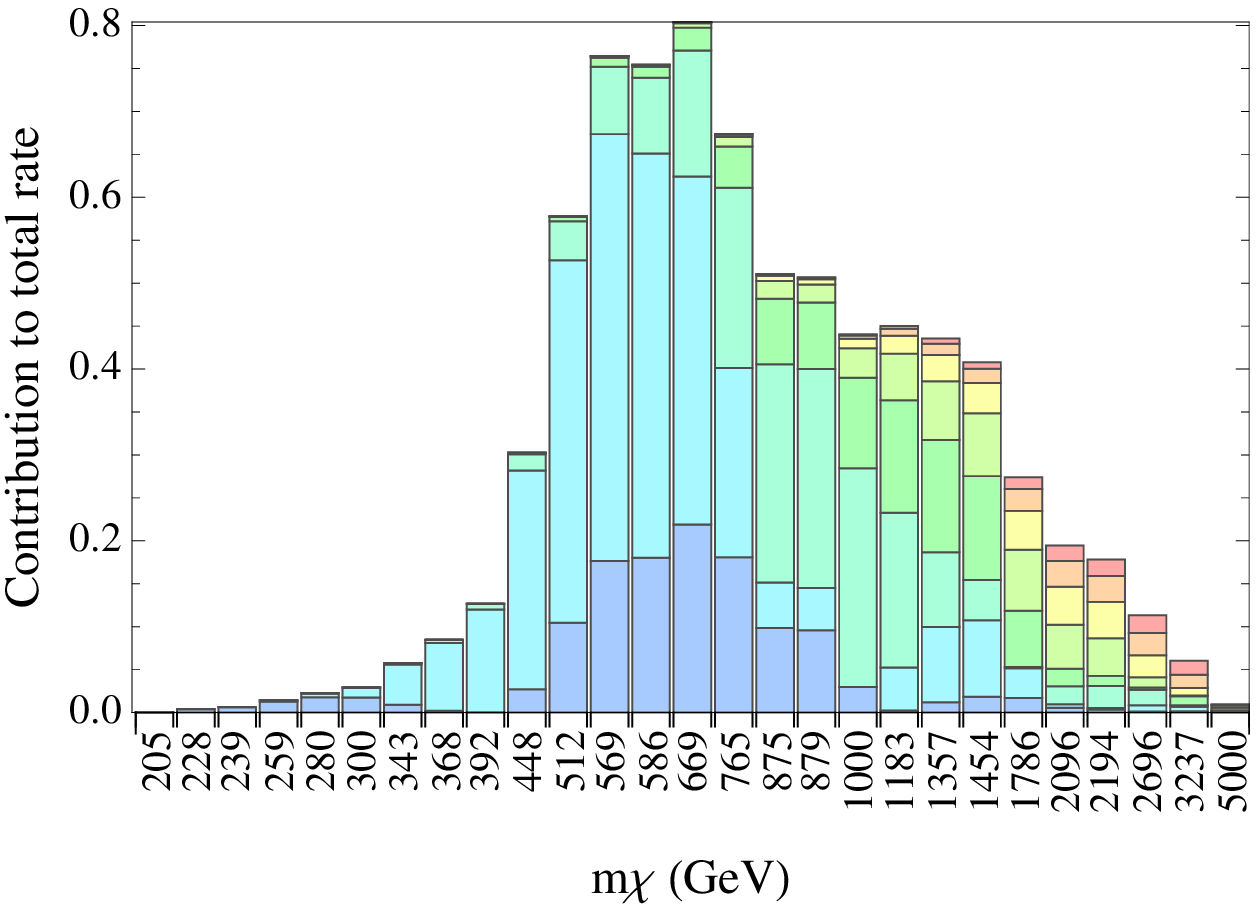}
\includegraphics[width=.53\textwidth]{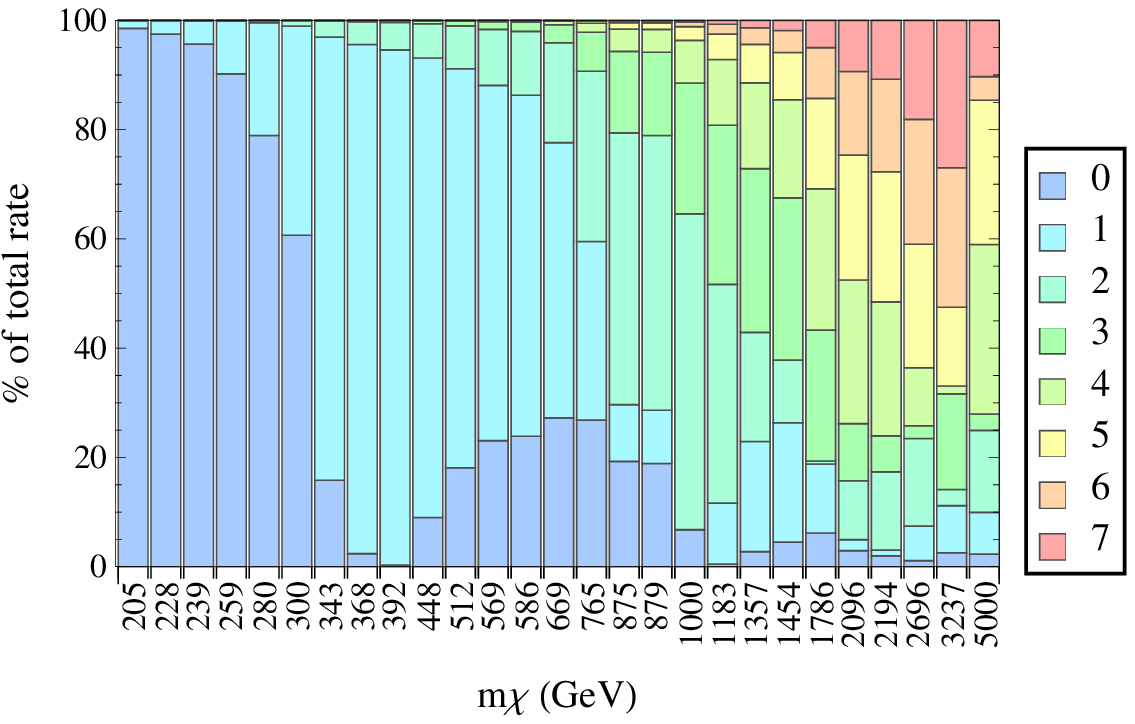}
\caption{Individual partial wave contributions for $m_\phi = 1\:\gev$ and $\Delta = 2\:\Mev$ varying $m_\chi$.}
\label{fig:pwcontributionsmchi2MeV}
\end{figure}

\begin{figure}[h]
\includegraphics[width=.47\textwidth]{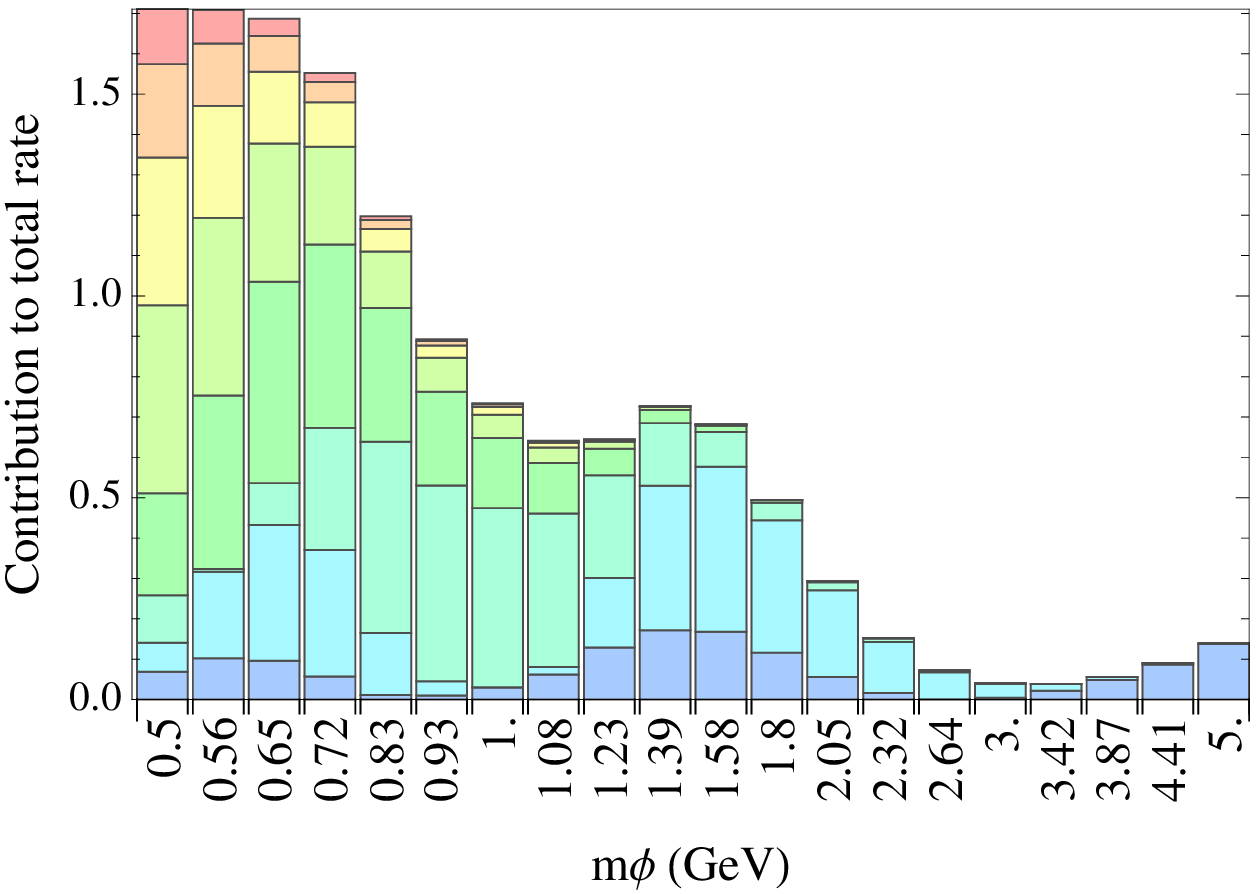}
\includegraphics[width=.53\textwidth]{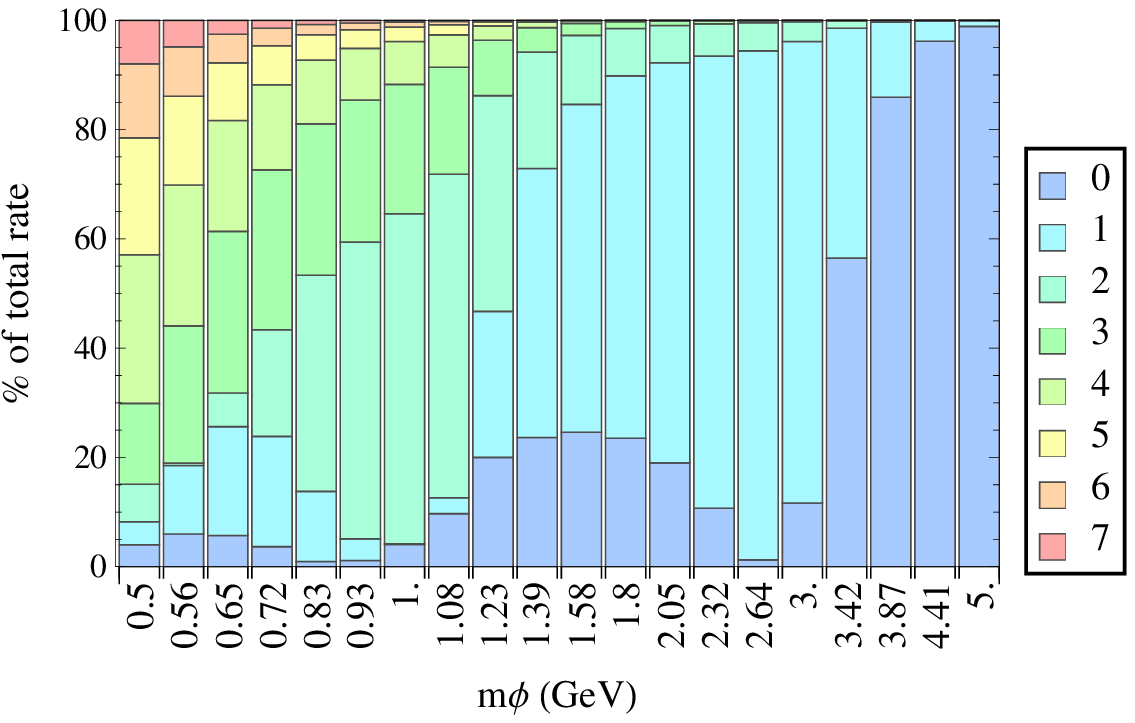}
\caption{Individual partial wave contributions for $m_\chi = 1\:\tev$ and $\Delta = 1\:\Mev$ varying $m_\phi$.}
\label{fig:pwcontributionsmphi1MeV}
\end{figure}

\begin{figure}[h]
\includegraphics[width=.47\textwidth]{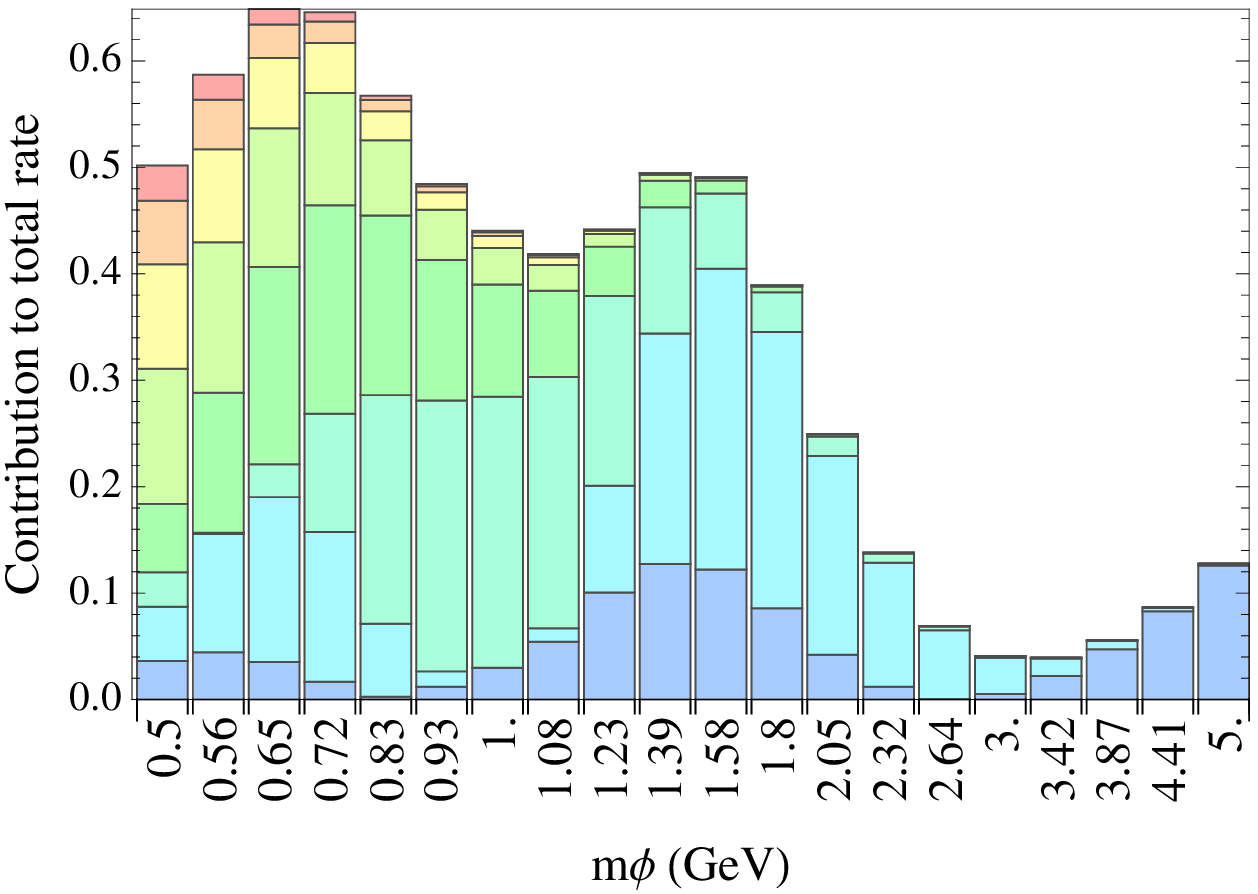}
\includegraphics[width=.53\textwidth]{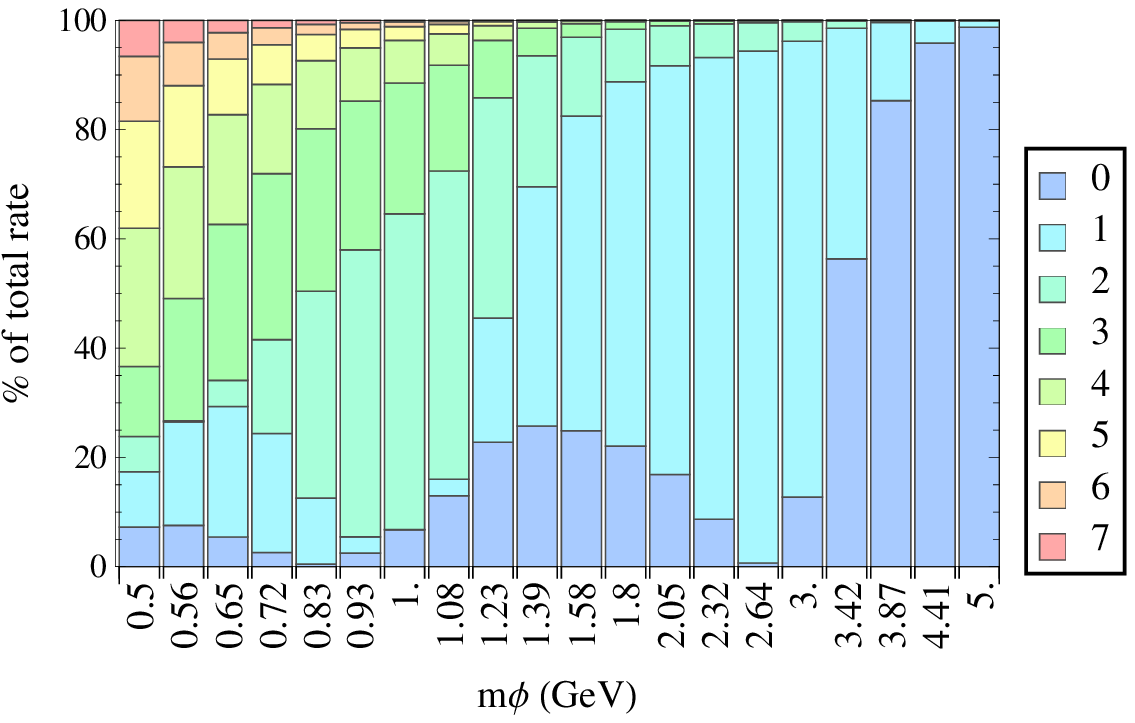}
\caption{Individual partial wave contributions for $m_\chi = 1\:\tev$ and $\Delta = 2\:\Mev$ varying $m_\phi$.}
\label{fig:pwcontributionsmphi2MeV}
\end{figure}

To better understand these rates, let us look at the partial wave composition of each rate. In figures \ref{fig:pwcontributionsmchi1MeV}, \ref{fig:pwcontributionsmchi2MeV}, \ref{fig:pwcontributionsmphi1MeV} and \ref{fig:pwcontributionsmphi2MeV} we show the absolute and fractional contribution to the total rate from each partial wave. We see that higher values of $m_\phi$ tend to be dominated by very low $l$ modes while the lower masses are more uniformly populated. This makes intuitive sense if we think of the Compton wavelength of $m_\phi$ as the characteristic radius for the angular momentum. Higher $m_\phi$ means the two WIMPs must get closer to upscatter and thus the system has lower angular momentum. The opposite case is true for the $m_\chi$ variations. Here the low $m_\chi$'s are dominated by low $l$ modes and the higher masses are evenly populated. This is because the lower the $m_\chi$, the higher the velocity needed to overcome the inelastic threshold. Thus the lower-mass WIMPs are all coming from the tail of the velocity distribution. This means that when they do upscatter, they are more likely to give up all their kinetic energy and the system is left without any angular momentum. A higher-mass WIMP at the same velocity will be capable of higher angular-momentum processes, as it can remain in a high $l$ state after the transition due to its relatively higher residual kinetic energy.

\subsection{Convergence of Sums}
\label{sec:rateconvergence}


\begin{figure}[t]
\includegraphics[width=.5\textwidth]{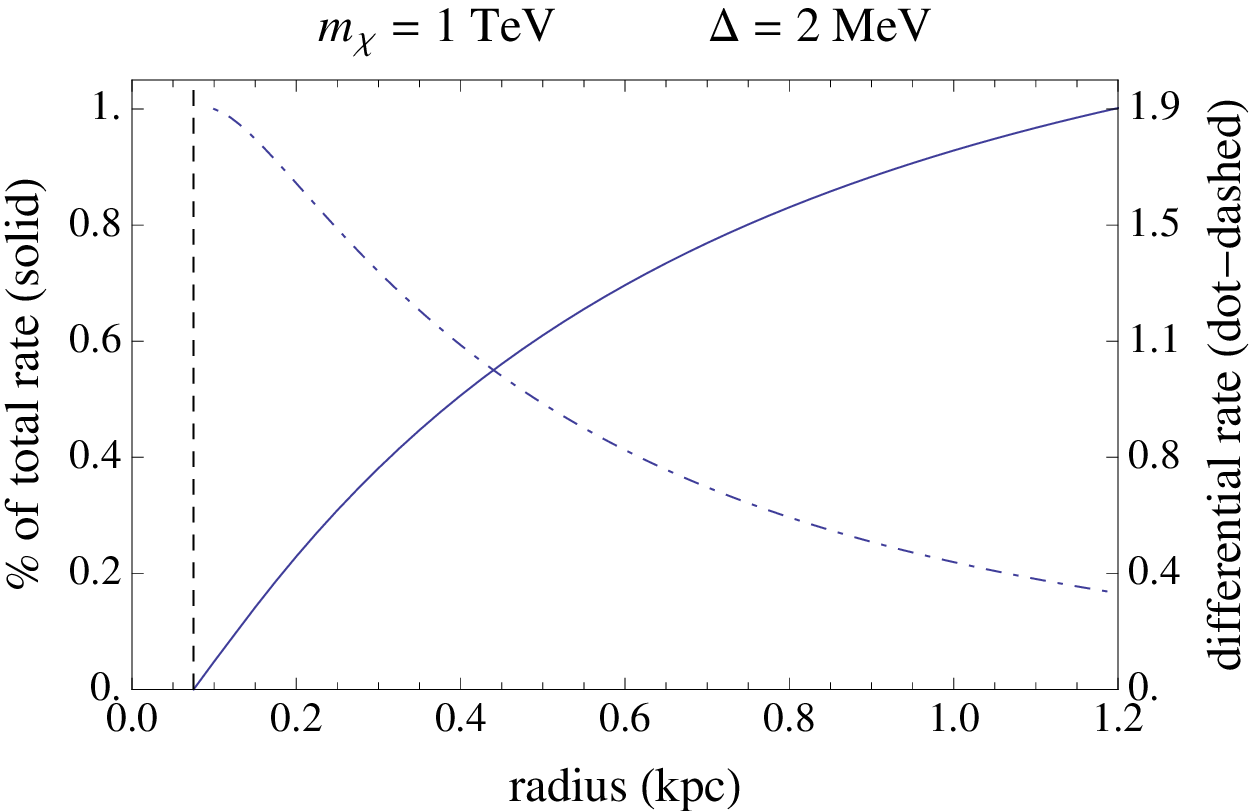}
\includegraphics[width=.5\textwidth]{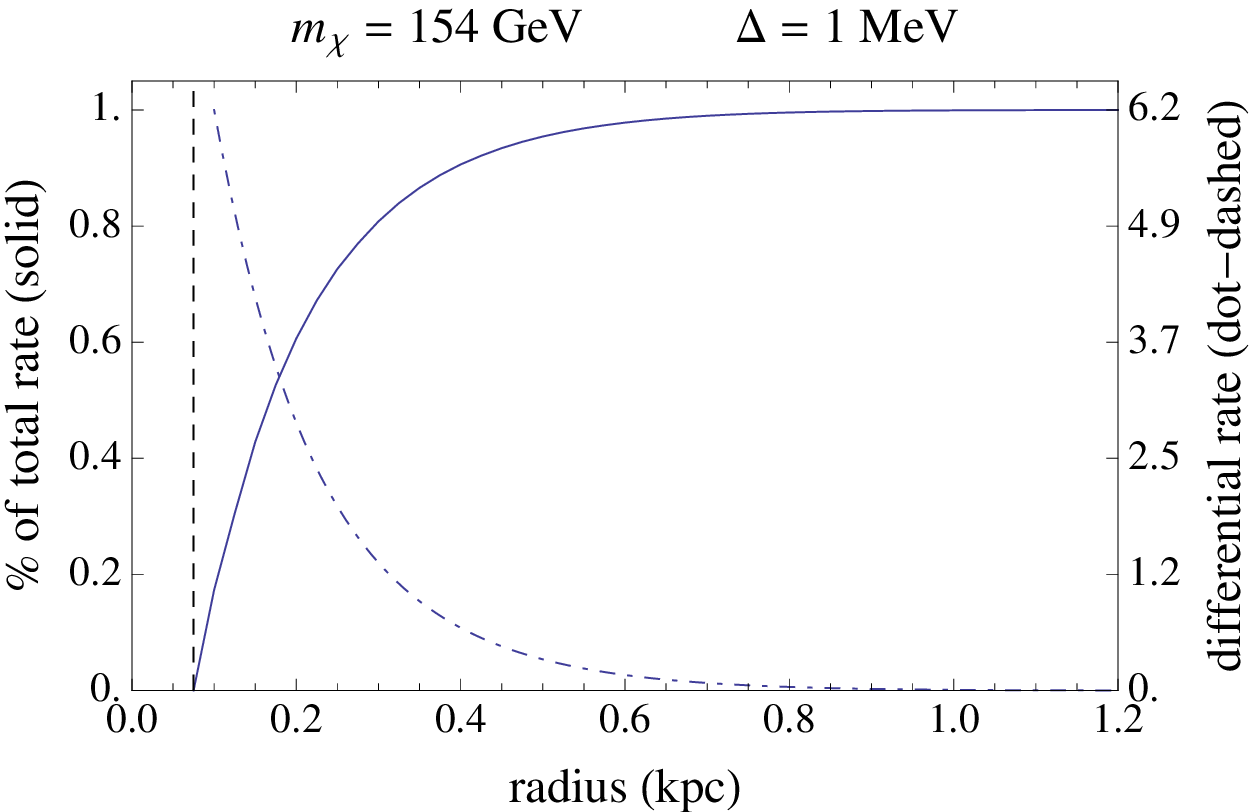}
\caption{Differential and total rates as a function of galactic radius. {\it (solid)} Percentage of total rate. {\it (dot-dashed)} Differential rate (in units of ${\rm kpc}^{-1}$) normalized by the total rate.}
\label{fig:totalratepercents}
\end{figure}

A concern we might have is that we may be significantly underestimating the partial wave sums with so few $l$ modes. To examine this, we have looked at how these saturate their ``total" as we increase the sum from $l=5$ to $l=7$. We find that when summing up to $l=5$ over $80\%$ of the points we considered are within $20\%$ of their values from summing up to $l=7$. With $l=6$, over $90\%$ of the parameter points are within $20\%$ and almost all of the points have reached at least $70\%$ of the $l=7$ total. Moreover, from section \ref{sec:partialwavecontributions} we can see that the rates with the most significant high $l$ contributions are either very high $m_\chi$ or very low $m_\phi$.  From this we conclude that in the most relevant regions of parameter space, the $l>7$ modes do not give significant contributions, and even where they do, we expect our results to be correct to $O(1)$. 

Because of the cuspiness in the profiles, another way the rates might be deceiving us is if they reach their total value in the first few hundred parsecs. For example, this could imply low mass WIMPs were garnering all their scatterings from questionably high velocities in the very center of the galaxy. In figure \ref{fig:totalratepercents} we plot the fraction of the total rate achieved as a function of galactic radius. We see that for the $1\:\tev$ WIMP with a splitting of $2\:\Mev$ the rate has reasonable contributions from all parts of the integral. The $154\:\gev$ WIMP with a $1\:\Mev$ splitting on the other hand picks up the majority of its rate in the first 300 pc. The accuracy of this rate relies upon the precise details of the RMS velocity profile, the escape velocity profile and the Einasto density profile in the very inner region of the galaxy. 

\begin{figure}[h]
\includegraphics[width=.5\textwidth]{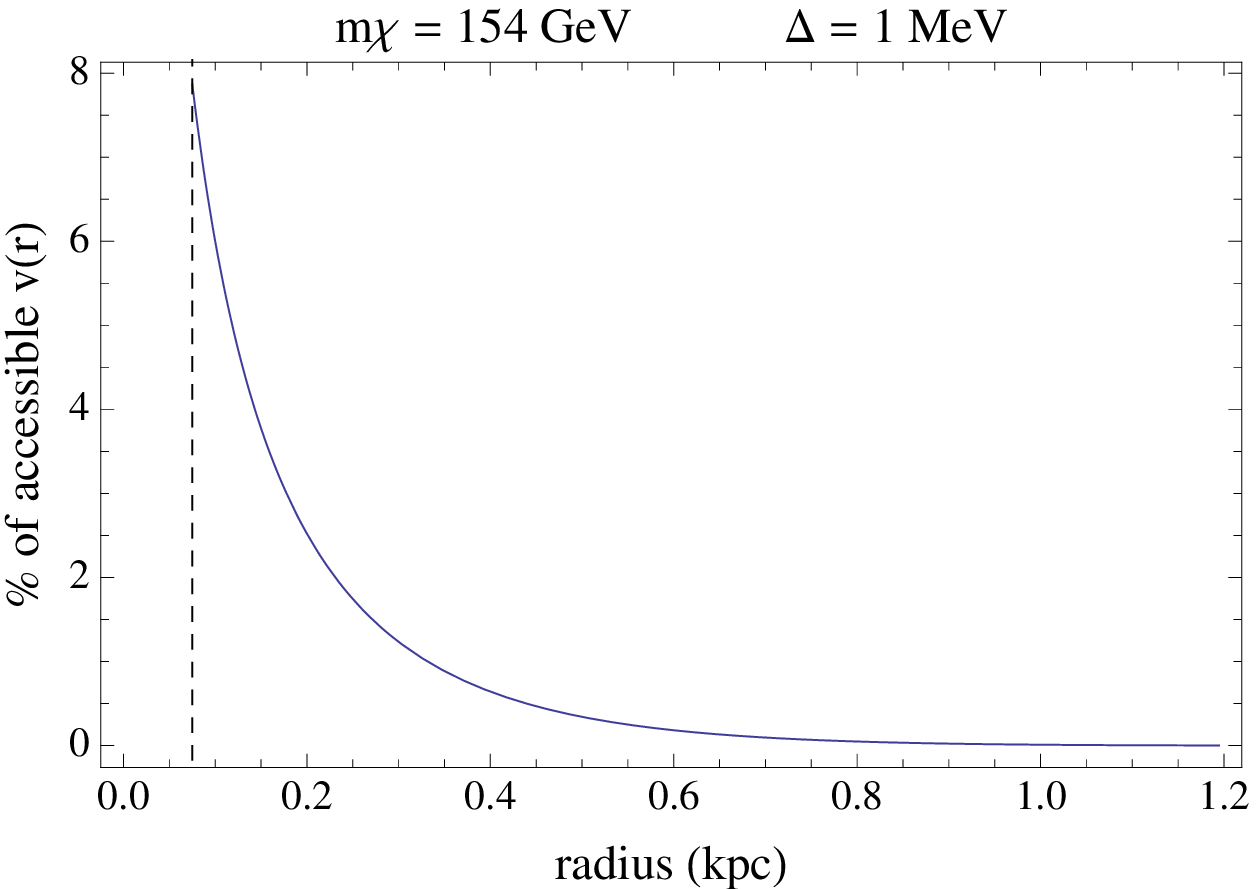}
\includegraphics[width=.5\textwidth]{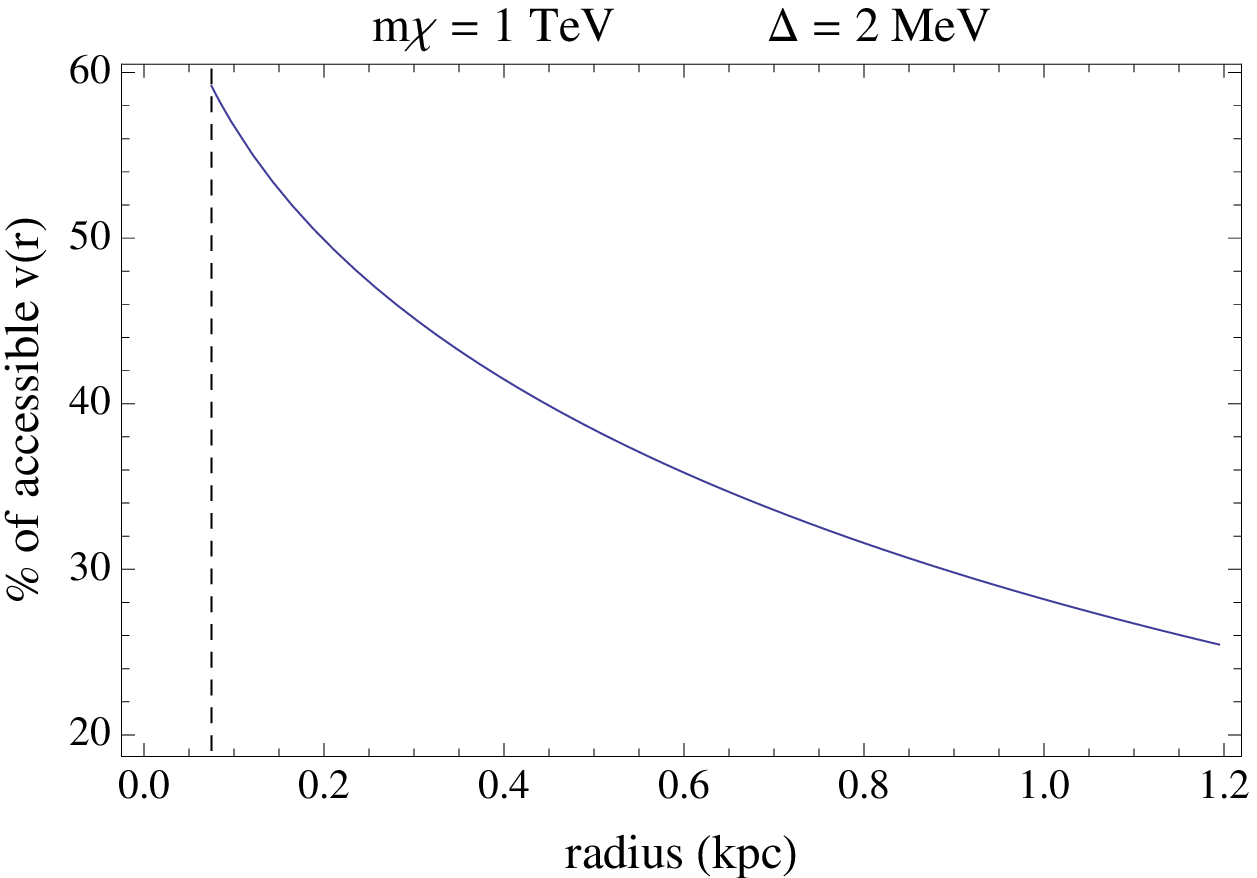}
\caption{Fraction of particles in the halo which are kinematically accessible for upscattering.}
\label{fig:allowedpairs}
\end{figure}

Another way to see this is to ask: at a given radius and for a given $\Delta/m_\chi$, what fraction of the velocity profile kinematically allows upscattering? We can plot this fraction as a function of radius to see if only the tail is contributing or if a significant portion of the WIMPs are contributing. We see in figure \ref{fig:allowedpairs} that, at best, only $4-6\%$ of the $154\:\gev$ WIMPs are upscattering while $30-60\%$ of the $1\:\tev$ WIMPs can upscatter. This reaffirms our conclusions in section \ref{sec:partialwavecontributions} that the low mass WIMPs (with only low $l$ contributions) are sampling from the tail, while higher mass WIMPs sample a broader range of particles. We expect that our results for higher mass WIMPs should be fairly robust.

\subsection{Profile Variations}
\label{sec:profilevariations}

For all of the rates shown so far the Aquarius A-1 profile values of $\alpha=0.17$ and $r_{-2}=15.79$ have been used. But since there is significant uncertainty in the profile parameters, let us consider the effects on our rates from variations of these parameters. In the following sections we will look at changes in the two Einasto parameters $\alpha$ and $r_{-2}$ as well as variations in the local escape velocity.

\subsubsection{Variations of $\alpha$}
\label{sec:alphavariations}
\begin{figure}[h]
\begin{center}
\includegraphics[scale=.75]{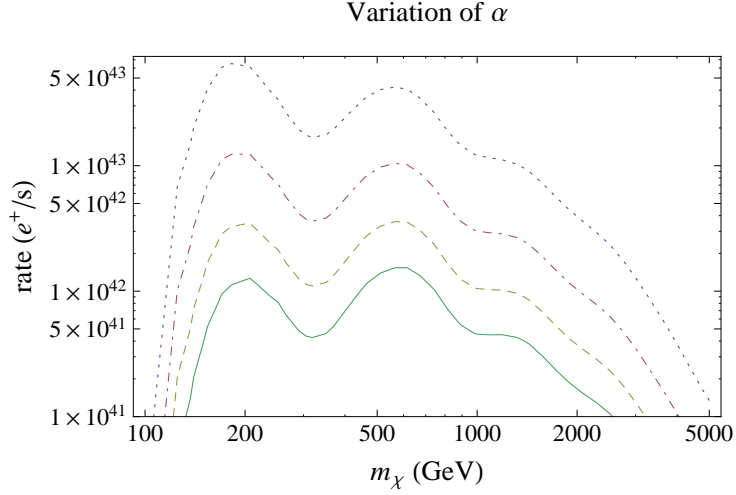}
\end{center}
\caption{Effects on rates of varying $\alpha$. {\it dotted}: $\alpha=0.05$, {\it dot-dashed}: $\alpha=0.1$, {\it dashed}: $\alpha=0.15$ and {\it solid}: $\alpha=0.2$}
\label{fig:alphavariations}
\end{figure}

In figure \ref{fig:alphavariations} we plot the effects on the rates of varying the Einasto profile parameter $\alpha$. We use as an example the scenario with $m_\phi=1\:\gev$ and $\Delta=1\:\Mev$ for various $m_\chi$. We vary $\alpha$ from 0.05 to 0.2 in steps of 0.05 while fixing $r_{-2}=15.79$, $v_c = 250\:\kms$ and $v_{loc} = 600\:\kms$. We see that a variation of $\alpha$ over this range can change the rates by an order of magnitude or more with the rate increasing as $\alpha$ decreases.

\subsubsection{Variations of $r_{-2}$}
\label{sec:r2variations}
\begin{figure}[h]
\begin{center}
\includegraphics[scale=.75]{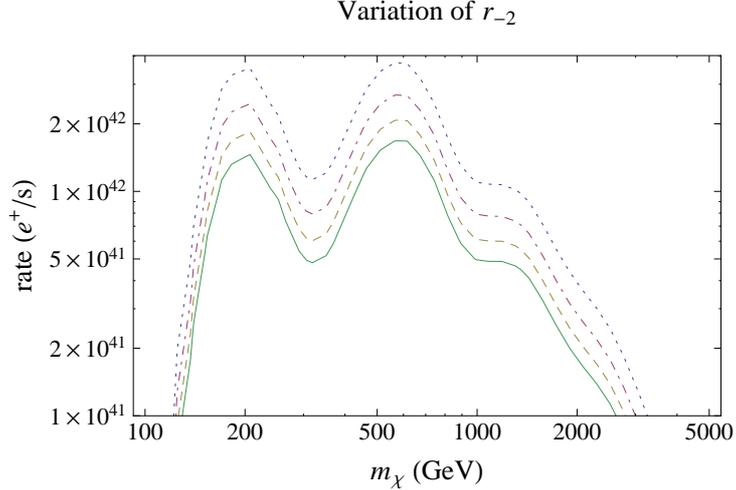}
\end{center}
\caption{Effects on rates of varying $r_{-2}$. {\it dotted}: $r_{-2}=12$, {\it dot-dashed}: $r_{-2}=15$, {\it dashed}: $r_{-2}=18$ and {\it solid}: $r_{-2}=21$}
\label{fig:r2variations}
\end{figure}

In figure \ref{fig:r2variations} we plot the effects on the rates of varying the Einasto profile parameter $r_{-2}$. We use the same example scenario of $m_\phi=1\:\gev$ and $\Delta=1\:\Mev$ while varying $m_\chi$. We fix $\alpha=0.17$, $v_c = 250\:\kms$ and $v_{loc} = 600\:\kms$ and then vary $r_{-2}$ from 12 to 21 in steps of 3. We see that at best variations in $r_{-2}$ can change the rates by about a factor of 5, with the lower $r_{-2}$ values giving higher rates.

\subsubsection{Variations of Local Escape Velocity}
\label{sec:vlocvariations}
\begin{figure}[h]
\begin{center}
\includegraphics[scale=.75]{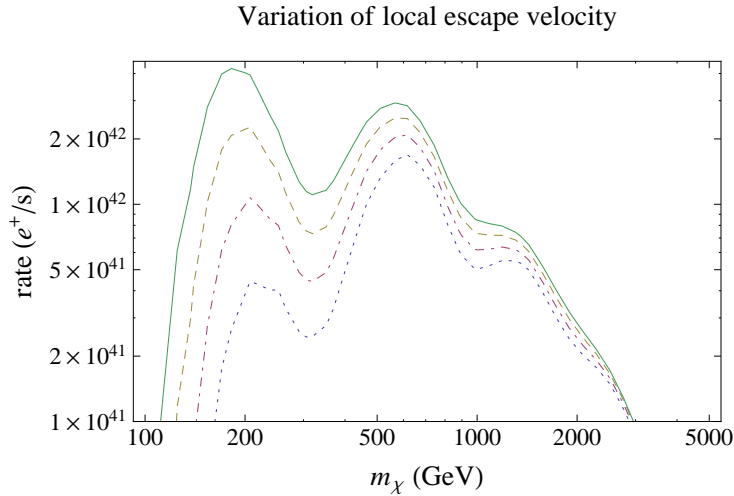}
\caption{Effects on rates of varying the local escape velocity. {\it dotted}: $v_{loc}=400\:\kms$, {\it dot-dashed}: $v_{loc}=500\:\kms$, {\it dashed}: $v_{loc}=600\:\kms$ and {\it solid}: $v_{loc}=700\:\kms$.}
\end{center}
\label{fig:vlocvariations}
\end{figure}

In figure \ref{fig:vlocvariations} we plot the effects on the rates of varying the local escape velocity (which in turn varies the escape velocity in the center of the galaxy). Again, we use the same example scenario of $m_\phi=1\:\gev$ and $\Delta=1\:\Mev$ while varying $m_\chi$. Here we fix $\alpha=0.17$, $r_{-2}=15.79$ and $v_c = 250\:\kms$.  We plot rates for local escape velocities of $400\:\kms$, $500\:\kms$, $600\:\kms$ and $700\:\kms$. As expected the rates go up for higher escape velocities, but the effect is mainly in the lower mass WIMPs. From $400\:\kms$ to $700\:\kms$ we see an overall enhancement of about a factor of 5 for $m\chi=500\:\gev$ and lower masses see an order of magnitude or more. In light of figure \ref{fig:totalratepercents}, these low $m_\chi$ enhancements are probably due to contributions in the innermost region of the galaxy. It is interesting to note that rates for $m\chi>1\:\tev$ are roughly independent of the local escape velocity. This is due to the fact that the higher mass WIMPs have high percentages of kinematically allowed pairs.

\section{Simulations with Baryons}
\label{sec:TWvariations}
Dark matter only simulations can probe high resolutions and short distances, but since we are most interested in the galactic center, where baryonic physics is important, the most reliable thing would be to employ simulations that also include baryonic physics. Recently, \cite{Tissera:2009cm} re-simulated many of the original, DM-only Aquarius runs while including the effects of baryons. They find that in the inner, baryon-dominated regions the halos become more concentrated. While the formation history plays a significant role in determining the characteristics of a galaxy's DM halo, the presence of baryons in the simulations seems to significantly increase the DM densities in the inner galactic region and this in turn provides significant increases to the $e^+ e^-$ production rates.  While the local DM densities and local velocity dispersions are similar to those used throughout this paper, their velocity dispersions have a much weaker dependence on radius.  In table \ref{table:simparameters} we give for reference various values from these re-simulated scenarios.

\begin{table}[h]
\begin{tabular}{|l|c|c|c|c|c|c|}
\hline
 & Aq-A-5 & Aq-B-5 & Aq-C-5 & Aq-D-5 & Aq-E-5 & Aq-F-5 \\ \hline
 $\alpha$ & 0.065 & 0.145 & 0.115 & 0.102 & 0.098 & 0.112 \\ \hline
 $r_{-2} \: (\rm{kpc} \, \rm{h}^{-1}) $ & 3.68 & 10.95 & 7.17 & 10.35 & 7.79 & 10.89 \\ \hline
 $\rm{log}\,\rho_{-2} \: (\rm{M}_\odot\, \rm{h}^2\, \rm{kpc}^{-3})$ & 7.81 & 6.59 & 7.28 & 6.85 & 6.99 & 6.62 \\ \hline
 Local Densities ($\gev \,\rm{cm}^{-3}$) & 0.51  & 0.26  & 0.57  & 0.43  & 0.35  & 0.28  \\ \hline
 Radial Scaling & $ r^{-0.162}$ & $ r^{-0.125}$ & $ r^{-0.144}$ & $ r^{-0.125}$ & $ r^{-0.125}$ & $ r^{-0.151}$ \\ \hline
 Local Dispersion ($\rm{km}\,\rm{s}^{-1}$)& 288.45  & 202.67  & 300.47  & 274.18  & 244.36  & 225.79 \\ \hline
\end{tabular}
\caption{Profile values from the DM plus baryons simulations.}
\label{table:simparameters}
\end{table}

\begin{figure}[h]
\includegraphics[width=.5\textwidth]{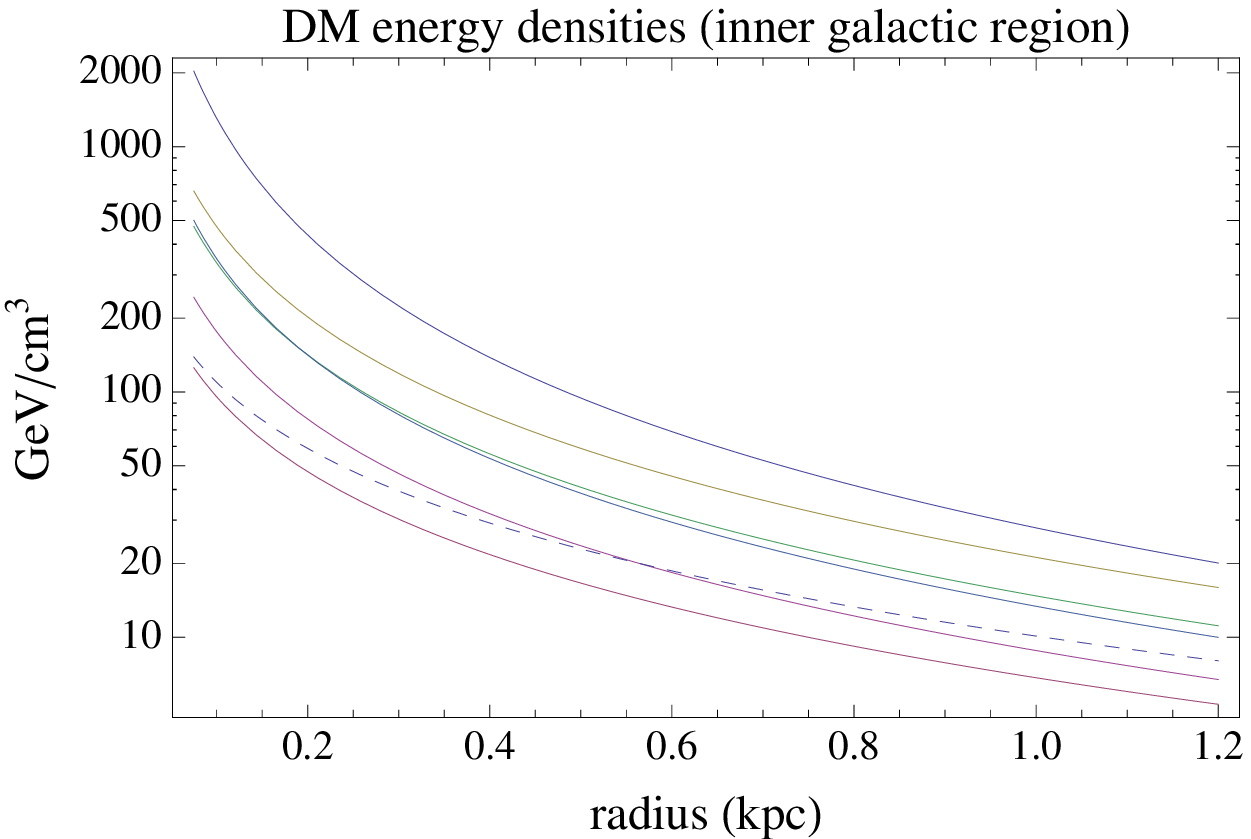}
\includegraphics[width=.5\textwidth]{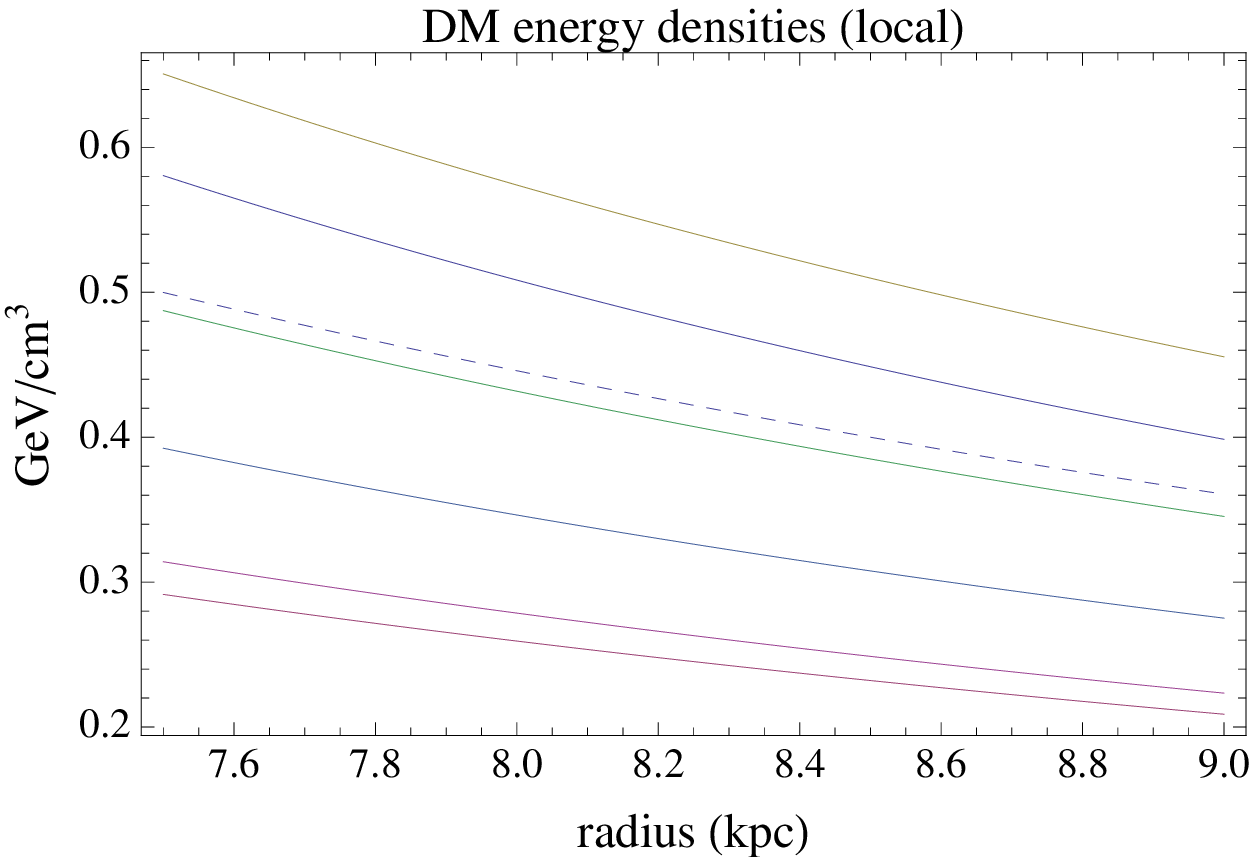}
\caption{Densities for a $100\:\gev$ WIMP in Aquarius runs re-simulated with baryons (see \cite{Tissera:2009cm}). {\it solid (from top to bottom)}: Aq-A-5, Aq-E-5, Aq-C-5, Aq-D-5, Aq-F-5, Aq-B-5. {\it dashed}: Aq-A-1 (DM-only).}
\label{fig:TWdensities}
\end{figure}

\begin{figure}[h]
\includegraphics[width=.5\textwidth]{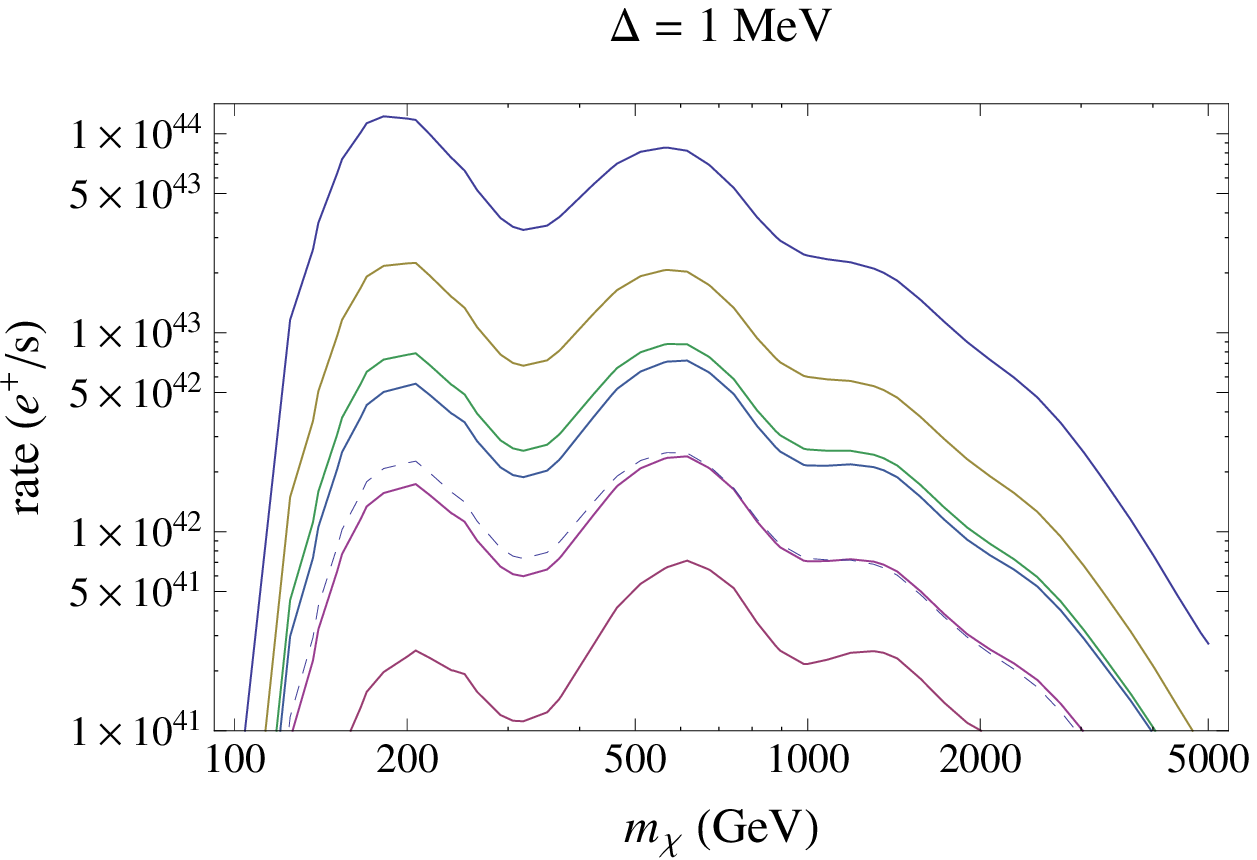}
\includegraphics[width=.5\textwidth]{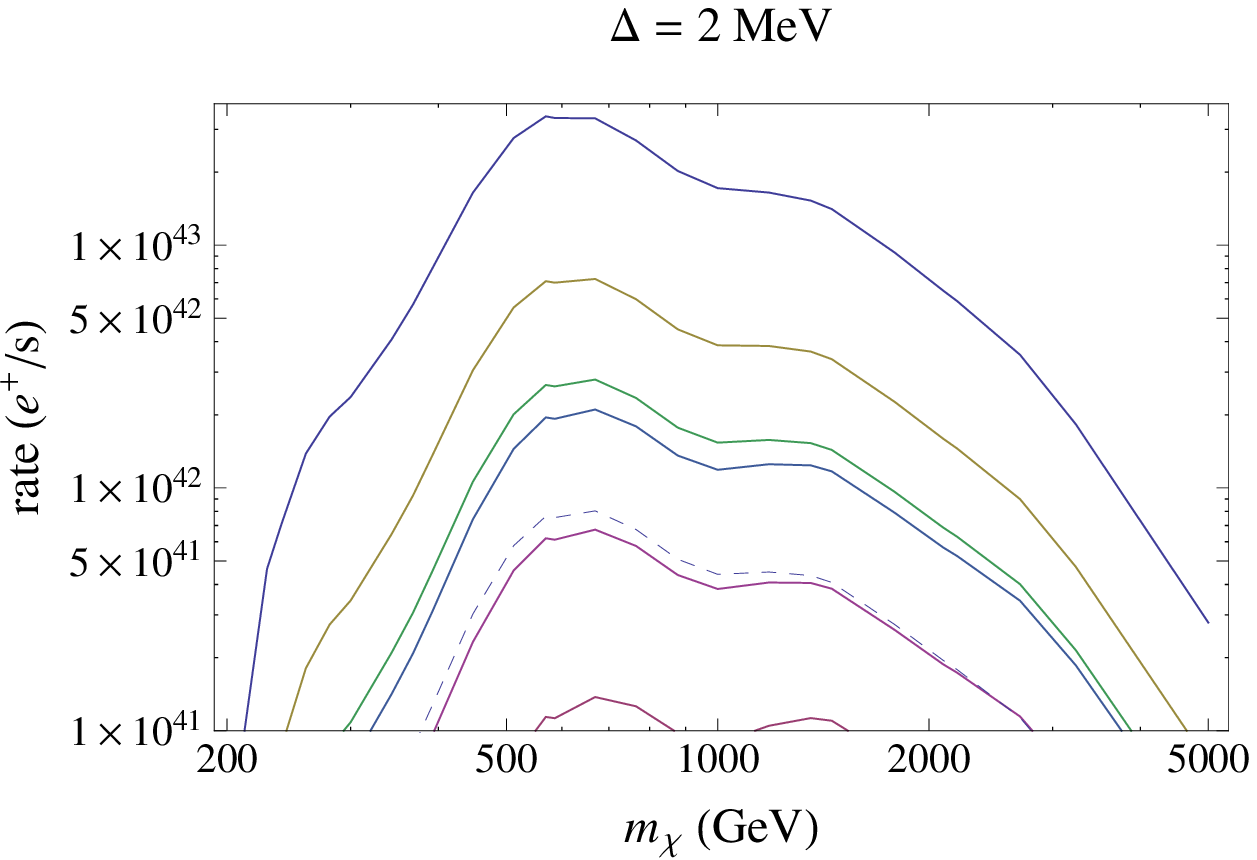}
\caption{Pair production rates for Aquarius runs re-simulated with baryons (see \cite{Tissera:2009cm}). {\it solid (from top to bottom)}: Aq-A-5, Aq-E-5, Aq-C-5, Aq-D-5, Aq-F-5, Aq-B-5. {\it dashed}: Aq-A-1 (DM-only). All lines assume $m_\phi= 1\:\gev$.}
\label{fig:TWmchivariations}
\end{figure}

\begin{figure}[h]
\includegraphics[width=.5\textwidth]{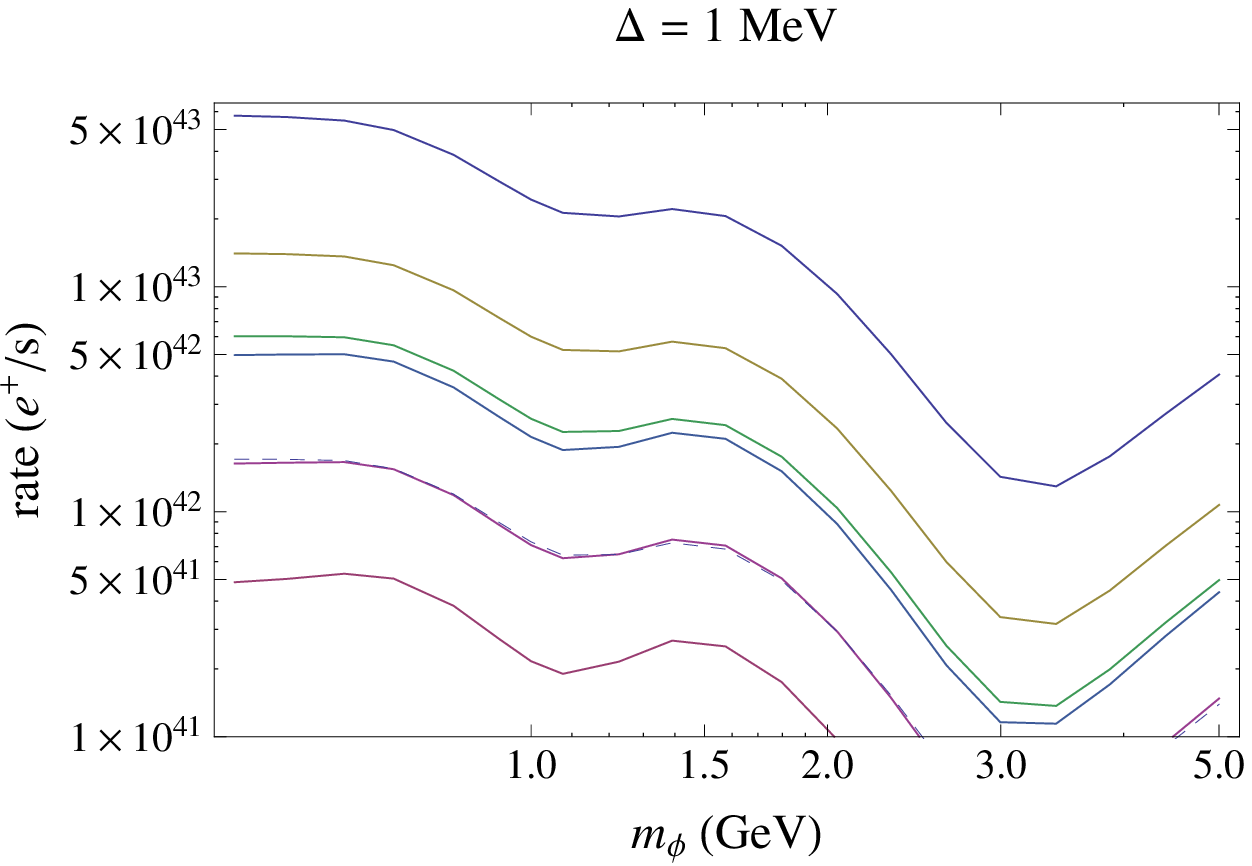}
\includegraphics[width=.5\textwidth]{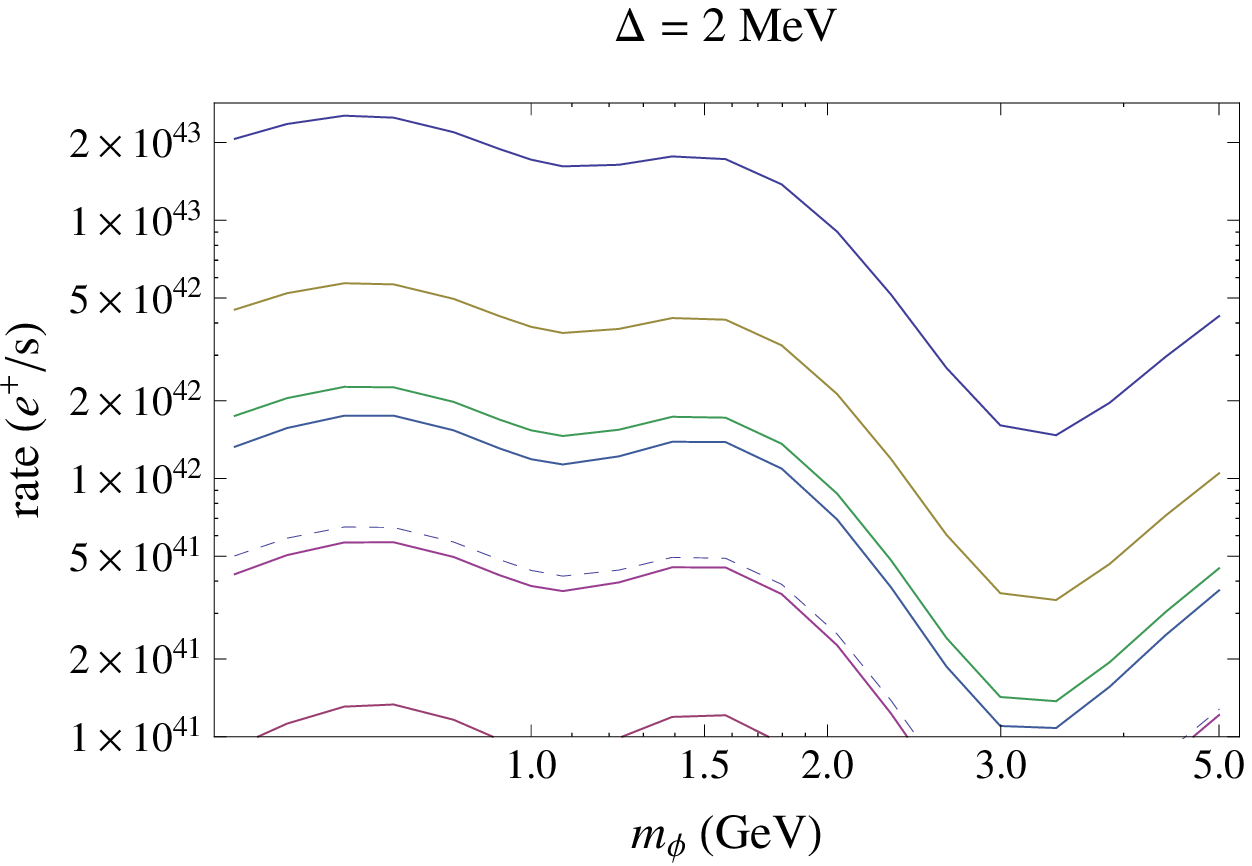}
\caption{Pair production rates for Aquarius runs re-simulated with baryons (see \cite{Tissera:2009cm}). {\it solid (from top to bottom)}: Aq-A-5, Aq-E-5, Aq-C-5, Aq-D-5, Aq-F-5, Aq-B-5. {\it dashed}: Aq-A-1 (DM-only). All lines assume $m_\chi= 1\:\tev$.}
\label{fig:TWmphivariations}
\end{figure}

We can then use the profile parameters from these simulations to calculate pair production rates. We present variations over $m_\phi$ and $m_\chi$ for both $\Delta = 1\:\Mev$ and $\Delta = 2\:\Mev$. Now though $\alpha$, $r_{-2}$, $\rho_{-2}$ and the velocity dispersion are all quantities determined from each individual simulation (see table \ref{table:simparameters}). As shown in figure \ref{fig:TWdensities} the DM densities in the inner part of the galaxy are generally higher than the A-1 DM-only simulation while their local densities can be higher or lower.  For most runs these increases in density give increases of roughly 3--40 to the rates, as shown in figures \ref{fig:TWmchivariations} and  \ref{fig:TWmphivariations}.

\subsection{Comparison with Previous Results}
\label{sec:clinecompare}
As a final note, we should make a direct comparison with the similar analysis of \cite{Chen:2009av}. In their study of the same problem, they concluded negatively that upscattering could provide the necessary rate to explain the INTEGRAL 511 keV signal absent the presence of populated metastable states. We propose two reasons why our analysis reached different conclusions from theirs, even when using the same Einasto profile parameters.  

First, they scanned a much broader range of parameters than we did but their results are quoted in terms of their dimensionless parameters. While experience has taught us to look for resonance regions in scans of $m_\phi$ and $m_\chi$, it is not immediately apparent what combinations of the dimensionless parameters will yield the same behaviors. Moreover, even having identified those, it is not obvious how fine a graining is required to find them. For these reasons we believe their scans may have simply missed the specific combinations of parameters we study and in particular resonance regions. 

We can compare directly with their results by noting that $\Gamma = \alpha_{d}^{2} \, m_\chi/\Delta$, $\eta = \alpha_d \, m_\phi/\Delta$ and $2\delta M = \Delta$.  Perhaps our best candidate point from the Aquarius A-1 profile is $m_\phi = 1\:\gev$, $m_\chi \simeq 600\:\gev$ and $\Delta = 1\:\Mev$ (recall that throughout this work we have set $\alpha_d = 1/100$). Translating this into their variables gives $\Gamma = 60$, $\eta = 10$ and $\delta M = 0.5\:\Mev$.  From the top-left group of their figure 5 they only have comparable values for $\Gamma=10$ and $\Gamma = 100$.  The two resonance regions we see along $\Delta = 1\:\Mev$ on the $m_\phi = 1\:\gev$ plot in figure \ref{fig:rates} would fall between these $\Gamma = 10$ and $\Gamma = 100$ plots.  The most direct comparison between our result and theirs might be for $m_\phi = 1\:\gev$, $m_\chi = 1000\:\gev$ and $\Delta = 1\:\Mev$.  From figure \ref{fig:rates} we can calculate $\rm{log}_{10}(R/R_{obs})\simeq -0.7$. This appears to be compatible with their equivalent point of $\Gamma = 100$, $\eta = 10$ and $\delta M = 0.5\:\Mev$.

Second, we believe the expression for the escape velocity used by \cite{Chen:2009av} is likely a significant underestimation. The expression used by \cite{Chen:2009av} is taken from the earlier \cite{Finkbeiner:2007kk}, and has a local  escape velocity of roughly $400\:\kms$, lower than most current estimates \cite{Smith:2006ym}.  From figure \ref{fig:vlocvariations} one can see that our rates drop by a factor of 5--10 for low mass WIMPs and 2--5 for higher mass WIMPs when we assume this local escape velocity. We believe that it is a combination of these two factors that lead us to different conclusions. 

More broadly, our inclusion of the recent results involving baryonic simulations further demonstrates that astrophysical effects can naturally boost the rates into the expected levels. This would likely affect the qualitative conclusions of  \cite{Chen:2009av} as well, but is outside the scope of their work and relies upon results that appeared subsequent to their paper.

\section{Conclusions}
\label{sec:conclusions}

Radiation from electron-positron annihilation in the center of the galaxy has been seen since the early 1970's: first by balloon-bourne experiments \cite{Johnson:1972, Johnson:1973, Haymes:1975} and later by satellites such as INTEGRAL. Today we understand there is a clear bulge plus disk morphology with the bulge having a full-width half-max of about $8^{\circ}$. The flux from the bulge component is estimated to be roughly $10^{43}$ pairs/s though it is unclear how much the disk is contributing to the bulge flux.  There are many proposed astrophysical sources in the center of the galaxy---pulsars, supernovae, LMXBs and microquasars to name a few---but no single source seems to have the right flux and shape.  All of these sources (except possibly LMXBs) are expected to have disk-like morphologies and total flux contributions on the order of $10^{43}$ pairs/s.  XDM scenarios naturally have a bulge-like shape due to the radial dependence of the number density.

In this work, we have performed a thorough numerical investigation into the plausibility of XDM scenarios explaining this bulge-shaped signal. To do this we first found the upscattering cross sections by solving the Schr\"odinger equation in the basis of partial waves for two two-particle states coupled through a Yukawa-type force. After integrating the cross sections over the relative velocity distribution to get the thermalized cross sections, we then find the rates of pair production by assuming an Einasto profile for the WIMP number density.  Due to numerical instabilities we were only able to calculate the first seven partial wave modes, but we find that in the range of reasonable model parameters this is sufficient. We expect uncertainties in the number density profile and the local galactic escape velocity to eclipse the uncertainty from higher $l$-mode contributions. 

Uncertainties in the calculation of electron-positron production rates in XDM scenarios come from both model parameters and astrophysical parameters.  While varying the Einasto profile parameters individually leads to monotonic changes in the rates, variations of the WIMP mass and force carrier mass are not so simple to predict.  We find that using the Aquarius A-1 Einasto values our pair production rates are on the order of $10^{41}$--$10^{42}$ pairs/s.  These rates could easily change by an order of magnitude or more via minor changes to the WIMP mass and profile parameters.  When using the Einasto parameters from Aquarius simulations including baryons we find rates that are generically of the order $10^{42}$--$10^{43}$ pairs/s and even as high as $10^{44}$ pairs/s. In light of this and the uncertainty in the bulge flux contribution from the disk, we believe XDM scenarios can provide a natural explanation of the anomalous electron-positron annihilation signal.

\bibliographystyle{JHEP}
\bibliography{xdm}

\end{document}